\documentclass[conference]{IEEEtran}
\IEEEoverridecommandlockouts
\usepackage{cite}
\usepackage{amsmath,amssymb,amsfonts}
\usepackage{algorithmic}
\usepackage{graphicx}
\usepackage{subfig}
\usepackage{caption}
\usepackage{diagbox}
\usepackage{booktabs}
\usepackage{textcomp}
\usepackage{xcolor}
\usepackage{url}
\def\BibTeX{{\rm B\kern-.05em{\sc i\kern-.025em b}\kern-.08em
    T\kern-.1667em\lower.7ex\hbox{E}\kern-.125emX}}
    
\newcounter{DaveCommentCounter}
\newcounter{LimingCommentCounter}
\usepackage{censor} 
\StopCensoring

\begin{document}


\title{Connecting Everyday Objects with the Metaverse:\\ 
A Unified Recognition Framework}

\author{\IEEEauthorblockN{Liming 
Xu\IEEEauthorrefmark{1}\thanks{\IEEEauthorrefmark{1}This work was completed while the first author was a Ph.D. student at University of Nottingham Ningbo China.}}
\IEEEauthorblockA{\textit{Department of Engineering} \\
\textit{University of Cambridge}\\
Cambridge, United Kingdom \\
lx249@cam.ac.uk}\\
\IEEEauthorblockN{Andrew P. French}
\IEEEauthorblockA{\textit{School of Computer Science} \\
\textit{University of Nottingham}\\
Nottingham, United Kingdom \\
andrew.p.french@nottingham.ac.uk}\\
\and
\IEEEauthorblockN{Dave Towey\IEEEauthorrefmark{2}\thanks{\IEEEauthorrefmark{2}Corresponding author.}}
\IEEEauthorblockA{\textit{School of Computer Science} \\
\textit{University of Nottingham Ningbo China}\\
Ningbo, China \\
dave.towey@nottingham.edu.cn}\\
\IEEEauthorblockN{Steve Benford}
\IEEEauthorblockA{\textit{School of Computer Science} \\
\textit{University of Nottingham}\\
Nottingham, United Kingdom \\
steve.benford@nottingham.ac.uk}
}

\maketitle



\begin{abstract}
The recent Facebook rebranding to Meta has drawn renewed attention to the metaverse.
Technology giants, amongst others, are increasingly embracing the vision and opportunities of a hybrid social experience that mixes physical and virtual interactions. 
As the metaverse gains in traction, it is expected that everyday objects may soon connect more closely with virtual elements. 
However, discovering this ``hidden'' virtual world will be a crucial first step to interacting with it in this new augmented world. 
In this paper, we address the problem of connecting physical objects with their virtual counterparts, especially through connections built upon visual markers. 
We propose a unified recognition framework that guides approaches to the metaverse access points.
We illustrate the use of the framework through experimental studies under different conditions, in which an interactive and visually attractive decoration pattern, an Artcode, is used as the approach to enable the connection. 
This paper will be of interest to, amongst others, researchers working in Interaction Design or Augmented Reality who are seeking techniques or guidelines for augmenting physical objects in an unobtrusive, complementary manner.
\end{abstract}

\begin{IEEEkeywords} 
Artcode, augmented reality, interaction, metaverse, visual marker
\end{IEEEkeywords}

\begin{figure*}[t]
    \centering
    \subfloat[Barcode\label{subfig:barcode}]{
        \includegraphics[width=0.13\linewidth]{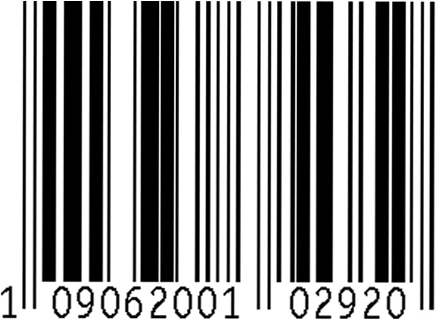}
        \hfill
    }
    \subfloat[QR code\label{subfig:qrcode}]{
        \includegraphics[width=0.095\linewidth]{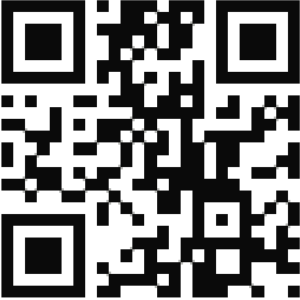}
        \hfill
    }
    \subfloat[Data matrix\label{subfig:data_matrix}]{
        \includegraphics[width=0.095\linewidth]{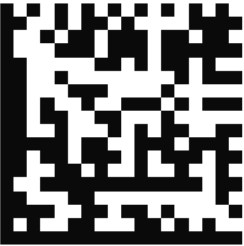}
        \hfill
    }
    \subfloat[Rohs' code\label{subfig:rohs_code}]{
        \includegraphics[width=0.09\linewidth]{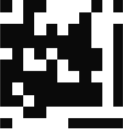}
        \hfill
    }
    \subfloat[ARTag\label{subfig:artag}]{%
        \includegraphics[width=0.095\linewidth]{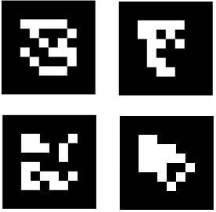}
        \hfill
    }
    \subfloat[ARToolkit\label{subfig:artoolkit}]{%
        \includegraphics[width=0.095\linewidth]{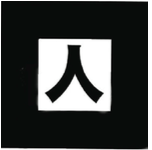}
         \hspace*{-0.75em}
    }
    \subfloat[ReacTIVision\label{subfig:reactivision}]{
        \includegraphics[width=0.0975\linewidth]{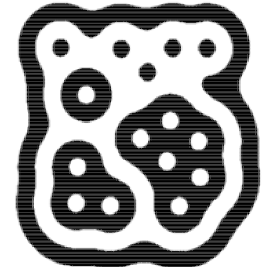}
         \hspace*{-0.5em}
    }
    \subfloat[D-touch\label{subfig:dtouch}]{
        \includegraphics[width=0.0925\linewidth]{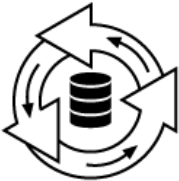}
         \hspace*{-0.75em}
    }
    \subfloat[Artcode\label{subfig:artcode}]{
        \includegraphics[width=0.09\linewidth]{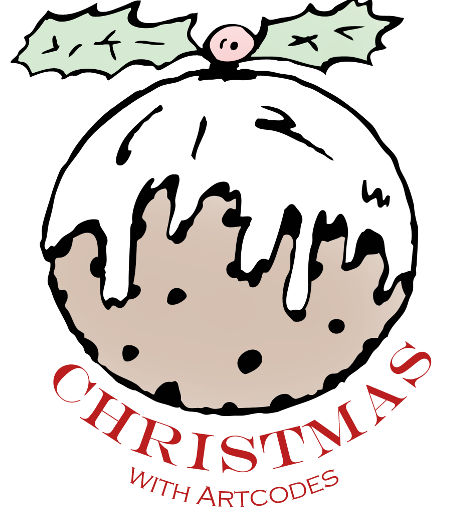}
    }
    \caption{Visual marker examples.}
    \label{fig:markerExamples} 
\end{figure*}

\section{Introduction
\label{sec:introduction}}


Attending events virtually has become a normalized part of our everyday life, due partly to the COVID-19 pandemic \cite{johnson2020virtual}. 
Increasingly, events are held online, or support attendance through avatars, on platforms such as Zoom, and Gather Town. 
This form of virtual engagement may well continue beyond COVID-19.
Moreover, Facebook's recent rebranding to Meta and Microsoft's announcement of launching into the metaverse strengthen the likelihood of this being part of our new normal \cite{waters2021microsoft}.
It is therefore reasonable to expect that our future will include a physical world even more augmented by a wide variety of virtual worlds. 
These virtual worlds may require unobtrusive and easy-to-use access points to a massive integrated network of virtual worlds or metaverse. 
Attainment of a fully-realized, immersive metaverse will require efforts and advances in multiple areas, including computer graphics, display hardware, and communication networks \cite{dionisio20133d}. 
In this paper, we address the issue of connections between the physical and the virtual worlds, proposing a conceptual framework for recognizing access points that may be hidden or camouflaged visual markers. 

The term ``metaverse'' was coined in 1992 by Neal Stephenson in his science-fiction novel \emph{Snow Crash} \cite{stephenson2003snow}, depicting a 3D virtual world where people can interact with each other, and with intelligent agents, through their avatars \cite{benford2021metaverse}. 
30 years later, and the development of metaverse is arguably still in its infancy, still with no generally accepted definition \cite{nevelsteen2018virtual, duan2021metaverse, benford2021metaverse}. 
The development framework of the metaverse, and its characteristics, have been studied in the literature. 
\blackout{Benford} \cite{benford2021metaverse}, for example, listed five metaverse properties: 
a virtual world;
a virtual reality; 
persistence; 
connection to the real world; and 
other people. 
In contrast to the industrial seven-layer metaverse value chain described by Radoff \cite{radoff2021metaverse}, 
Duan et al. \cite{duan2021metaverse} proposed a three-layer metaverse development architecture,
representing the physical world, interaction, and the virtual world. 
In spite of the lack of consensus on definition, 
there does appear to be general agreement that three basic metaverse properties are: 
(i) a physical world; 
(ii) a virtual world; and 
(iii) the connection between these two worlds. 

Although various devices have been designed for accessing virtual elements or virtual worlds, a map showing the presence of access points to these virtual worlds would guide the connection (and potentially enhance the experience). 
If this could be provided in an explicit and straightforward manner, for example, through an annotation indicating the presence of such entrances to virtual worlds, then even better!
In contexts requiring aesthetic-awareness, such at art galleries, implicit markers integrated into a part of the environment (such as in the surface pattern of an object) may be more appealing. 
In other environments, like in a corridor or hallway, both implicit and explicit visual markers may be acceptable. 
In this paper, we report on the use of such surface visual markers for connecting everyday objects with digital materials 
--- 
such as digital footprints, a virtual world, or a metaverse. 
We propose a unified recognition framework (URF) for bridging the physical and virtual worlds through visual decorations. 

The main contributions of this paper are threefold, summarised as follows:
\begin{itemize}
    \item We report on the use of visual markers as clues to prompt interaction with virtual worlds.
    \item We generalize a URF for identifying the presence of access points in public spaces. 
    \item We report on experimental studies conducted using one type of visual marker (Artcodes \cite{meese2013codes, benford2015augmenting}), illustrating how the proposed URF works. 
\end{itemize} 

The rest of this paper is organized as follows. 
Section \ref{sec:relatedWork} briefly reviews the related work on visual markers in augmented (AR) and virtual reality (VR). 
Section \ref{sec:framework} introduces the URF and the preliminaries pertaining to this work. 
Section \ref{sec:experiment} describes experimental studies evaluating the use of Artcodes as access points to virtual elements. 
Section \ref{sec:discussion} includes discussion of the implications of this study.
Finally, Section \ref{sec:conclusion} concludes this paper and describes future work. 

\begin{figure*}[t]
    \centering
    \includegraphics[width=\textwidth]{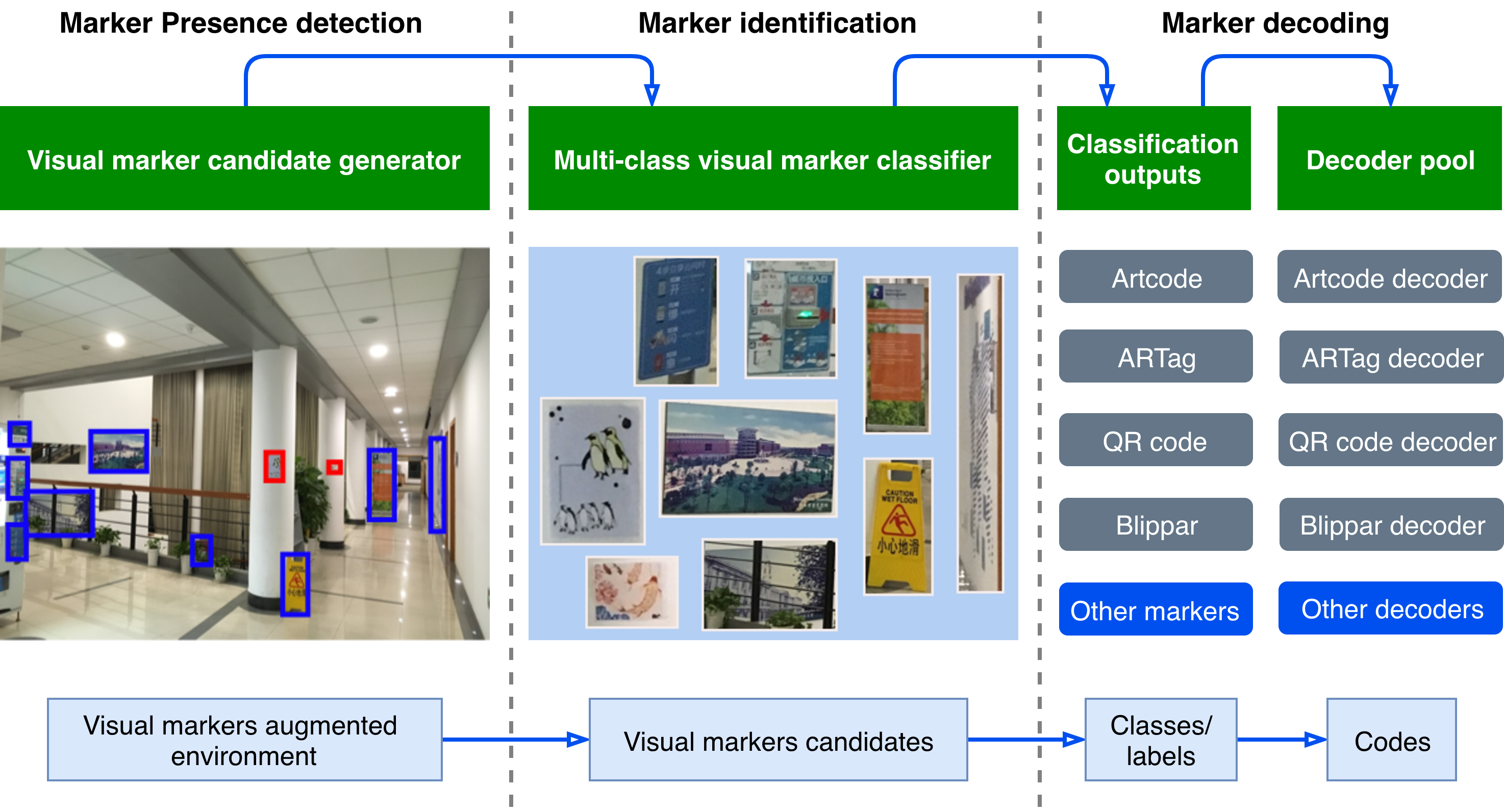}
    \caption{A unified recognition framework (URF) for visual markers.}
    \label{fig:framework}
\end{figure*}

\section{Related work on visual markers
\label{sec:relatedWork}}

A variety of visual markers (see examples in Figure~\ref{fig:markerExamples}), both human-readable and not, have been proposed \cite{costanza2009designable, meese2013codes}, with two of the most well-known being barcodes \cite{woodland1952classifying} (Figure~\ref{subfig:barcode}) and QR codes (Quick Response codes, Figure~\ref{subfig:qrcode}) \cite{qrcode2015}.
The barcode was among the earliest methods of representing data in a visual, machine-readable form, initially patented in 1952 \cite{meese2013codes}. 
While barcodes mainly appear in the retail sector, QR codes have become a ubiquitous feature \cite{meese2013codes}.  
Barcodes and QR codes were designed to be reliably read by machines, with no error occurring when they are scanned.
However, this reliability comes at a cost of limited aesthetics: 
Neither are visually meaningful to humans, and it can be difficult to distinguish different codes though visual inspection alone.

Many other visual marker systems have similar characteristics to barcodes and QR codes, often with their information being encoded within a matrix of black and white dots, and usually with some form of error detection and correction mechanisms. 
Examples of such marker systems include the Data Matrix \cite{datamatrix2006} (Figure~\ref{subfig:data_matrix}) and the Rohs visual code \cite{rohs2005conceptual} (Figure~\ref{subfig:rohs_code}).
While these visual markers are effective for encoding data, they were not intended for camera pose estimation and calibration, and are thus not appropriate for use as fiducials in AR systems
---
a fiducial is a type of marker mounted within an environment to enable estimation of the relative pose between the camera and object. 
Some example fiducial systems are: 
ARTag (Figure~\ref{subfig:artag}) \cite{fiala2005artag};  
ARToolkit (Figure~\ref{subfig:artoolkit}) \cite{kato1999marker}; and
reacTIVison (Figure~\ref{subfig:reactivision}) \cite{bencina2005improved}. 
ARTag markers employ a square border for marker localization, connectivity and perimeter analysis. 
They have a large library of patterns inside the border and use edge-detection approaches to achieve reliability \cite{fiala2005artag}.
ARToolkit markers consist of a thick square black border with a variety of patterns in the interior
---
the black outline allows for marker localisation and 
\emph{homography}\footnote{An isomorphism in projective spaces that is used to calibrate camera pose.} 
calculation.
The reacTIVision markers are automatically generated by fiducial recognition engines such as Amoeba and D-touch \cite{costanza2003d}: 
They have compact geometry and offer a limited space for users to adjust their aesthetic aspects \cite{bencina2005improved}.

The visual appearance of marker systems that rely on geometrical features for localization and encoding is strongly constrained. 
In the majority of cases, the shape (the geometry) of the markers is automatically generated, allowing little freedom of design. 
In contrast, another type of visual markers, such as D-touch and its variant Artcodes \cite{meese2013codes, xu2019artcode}, offer much more flexibility in geometrical form, both for the outline shape and the interior elements. 
D-touch encodes information through the topological structure of the markers 
--- 
the adjacency information of connected components, represented in a region adjacency tree \cite{costanza2003region}.
This supports users' creation of their own readable markers that are both aesthetic and meaningful \cite{costanza2009designable}.
Artcode implements and extends the D-touch approach, refining their drawing rules, and introducing human-meaningful (but machine-irrelevant) embellishments and aesthetic style guidelines. 
The Artcode approach provides the creative freedom to produce visually appealing \emph{and} machine-readable markers (patterns) that are meaningful to humans, and that resemble free-form images. 

In addition to these visual marker technologies based on geometry or topology, conventional image recognition technologies have also been employed to relate information to a much wider variety of images. 
Blippar \cite{blippar2021} and Google Lens \cite{google2021lens}, for example, make use of image recognition techniques to embed data into images. 
However, because these techniques often use neural networks and vector matching for encoding and decoding information, it is challenging (or impossible) to explain and interpret how the system works to non-technical designers or users. 
More recently, new systems that use deep-generative networks to automatically generate markers have been proposed, including learnable visual markers \cite{grinchuk2016learnable}, E2ETag \cite{brennan2021e2etag} and DeepFormableTag \cite{yaldiz2021deepformabletag}.
\section{Unified Recognition Framework (URF)
\label{sec:framework}}

As AR and metaverse applications become more pervasive, we will live in a world with dispersed access points to connect with virtual elements.
There will be an increasing number of entrances to these elements within our surrounding environment,
through a variety of virtual markers, both visible and ``hidden''.
Identifying the probable existence of these entrances will be the first step to triggering the follow-up interaction. 
Considering the many types of entrance that may co-exist, a unified recognition framework (URF) will be needed. 
In this section, we present such a conceptual URF for general visual marker presence recognition and identification.

Given the number of extant visual markers, both in academia and in industry, and the high likelihood of many more systems emerging in the future, attempting to explicitly include all in this URF would be unrealistic.
We therefore only include a selection of some typical markers to show the basic URF components.
The left part of Figure~\ref{fig:framework} shows a common scene, an indoor area of a building with various visual markers (highlighted in the picture).
Not all of the annotated objects are readable
---
some are explicitly-placed readable Artcodes (in red boxes), while others (in blue boxes) are commonplace objects that could be enhanced as visual markers.

As shown in Figure \ref{fig:framework}, the URF involves three stages: 
marker presence detection; 
marker identification;
and marker decoding. 
The \emph{detection} stage involves detecting visual markers in the surrounding environment. 
Given the scenario in the left part of Figure \ref{fig:framework}), for example, this stage would detect the possible presence of visual markers using image processing and computer vision techniques, and would output a set of localized candidate visual markers. 
This output set is then passed to the \emph{identification} stage (the middle of Figure \ref{fig:framework}) to determine if they \emph{are} markers, and, if so, what class of markers they belong to (Artcodes, QR codes, Blippar images, etc.). 
A key component of the identification stage is a \emph{multi-label classifier} that accepts the candidate markers, and outputs their corresponding classes or labels. 
The final stage is the \emph{decoding}, which includes a \emph{decoder pool} from within which the corresponding decoder identifies and decodes the embedded message in the visual marker. 

Once the data (codes) carried by the visual marker are identified, the connected visual information (labelled by the visual marker) can be triggered. 
In this URF, visual marker detection and identification are two independent stages, but in reality, these two things are often done together.
Although the URF is a conceptual framework, describing the essential components and a feasible pipeline to bridge the physical and virtual worlds, 
the concrete implementation may differ from one scenario to another. 
A possible URF implementation may be an \emph{all-in-one} brokering system that recognizes the presence of all (or most) of the visible or hidden visual markers, then calls the corresponding decoders or identifiers, and then steps into the embedded virtual worlds. 

The next section presents experimental studies examining discovery of the presence of visual markers using a concrete marker system, Artcode \cite{meese2013codes, xu2017recognizing}.

\section{Experimental studies
\label{sec:experiment}}

The URF proposed in the last section includes the two primary elements: 
visual marker discovery and identification, with discovery of the markers being a \emph{prerequisite} to the follow-up identification. 
Moreover, providing hints and clues to the location of (camouflaged) access points to virtual worlds may encourage people to explore those connections, thus creating new interaction opportunities. 
Given the importance of visual marker discovery in the URF pipeline, we conducted two case studies into how digital clues can be provided to guide users with devices (such as AR headsets) to approach the object and enter the metaverse. 
Artcodes, which are both meaningful to humans, and readable by scanners, were selected as the marker system.

\subsection{The Artcode approach
\label{subsec:artcode}}

Artcodes\footnote{https://www.artcodes.co.uk/} 
are human-designable topological visual markers, developed based on the D-touch system \cite{costanza2009designable}.
By incorporating additional drawing constraints and aesthetic embellishments, 
Artcodes enable more visually pleasing and interactive patterns than d-touch \cite{meese2013codes}.
Figures \ref{subfig:dtouch} and \ref{subfig:artcode} show examples of d-touch and Artcode markers. 
A valid Artcode consists of two parts: 
a recognizable foreground (the food image in Figure \ref{subfig:artcode}); and 
some image-based background (the text in Figure \ref{subfig:artcode}).
The foreground is intended for reading by machines, but the background can be designed for human consumption. 
Artcodes can be beautiful, interactive motifs that can decorate the surface of everyday objects without impacting the aesthetics of the object in the way that QR codes would. 

Because of their unobtrusive and non-obvious properties, the presence of an Artcode is not usually obvious:
Close inspection may be needed to discover an Artcode when there are no visual clues. 
Detection of Artcodes through their general visual features, identifying their probable locations by means of a \emph{heat map}, is therefore a meaningful approach.
Given the space limitations of this article, interested readers are referred to the literature for more information about Artcodes, including their design, detection, and identification \cite{meese2013codes, costanza2003d, xu2017recognizing, xu2019artcode, MRArtcodesJSS2021}.

\subsection{Experimental setting}\label{subsec:setting}
We conducted experiments to explore Artcode detection in an environment, and deliver clues to guide the subsequent interaction. 
We assumed a realistic interaction scenario, in which users may wear or carry devices in a physical space, standing far away from the Artcodes:
When they discover the presence of an Artcode, they can follow clues to approach the target for further interaction. 
Rather than fully simulating this scenario, we simplified it while maintaining its core characteristics: 
Users gain increasing amounts of details as they approach the target.

\begin{figure} [t]
    \centering
    \addtocounter{subfigure}{-16}
    \subfloat{
        \includegraphics[width=0.225\linewidth]{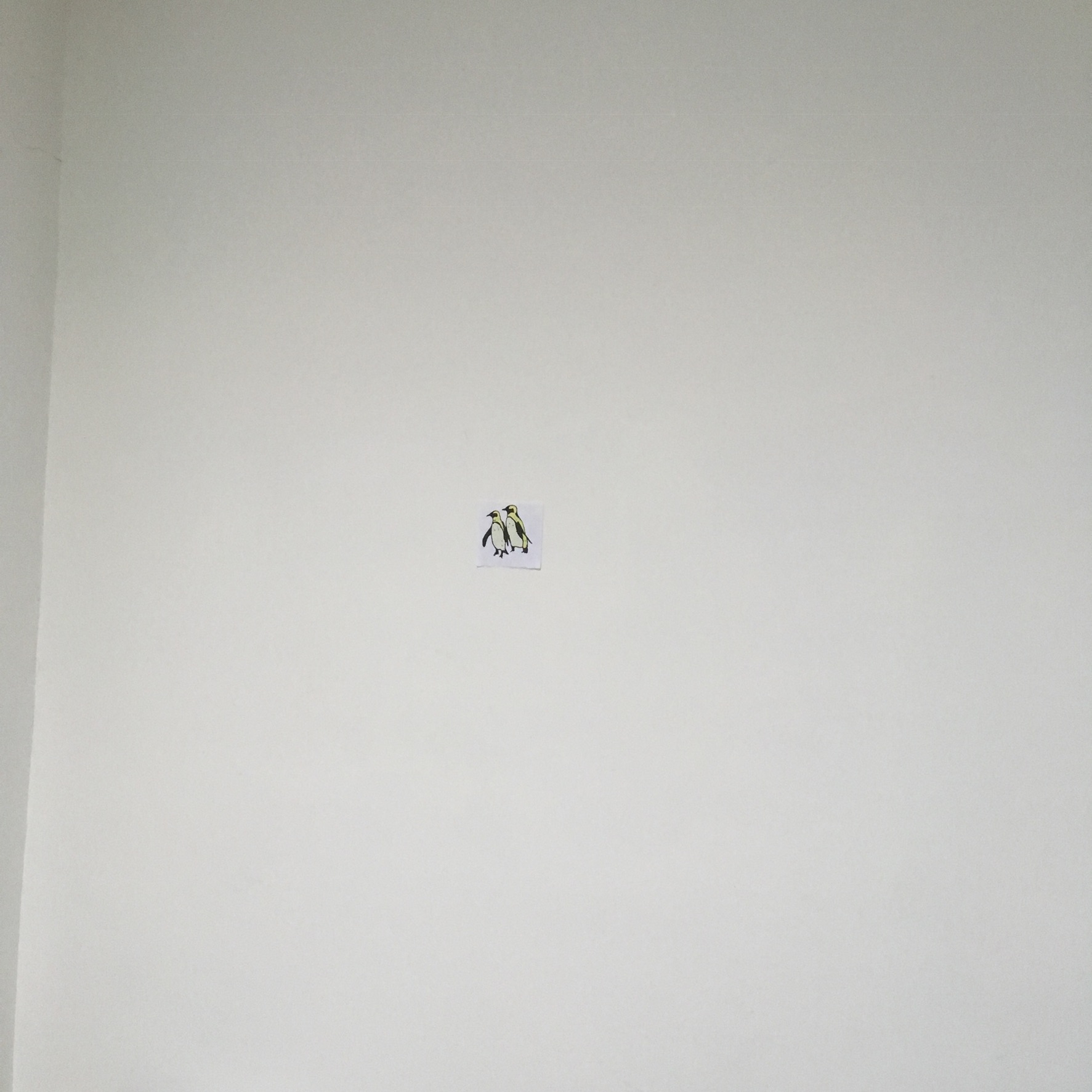}
    }
    \subfloat{
        \includegraphics[width=0.225\linewidth]{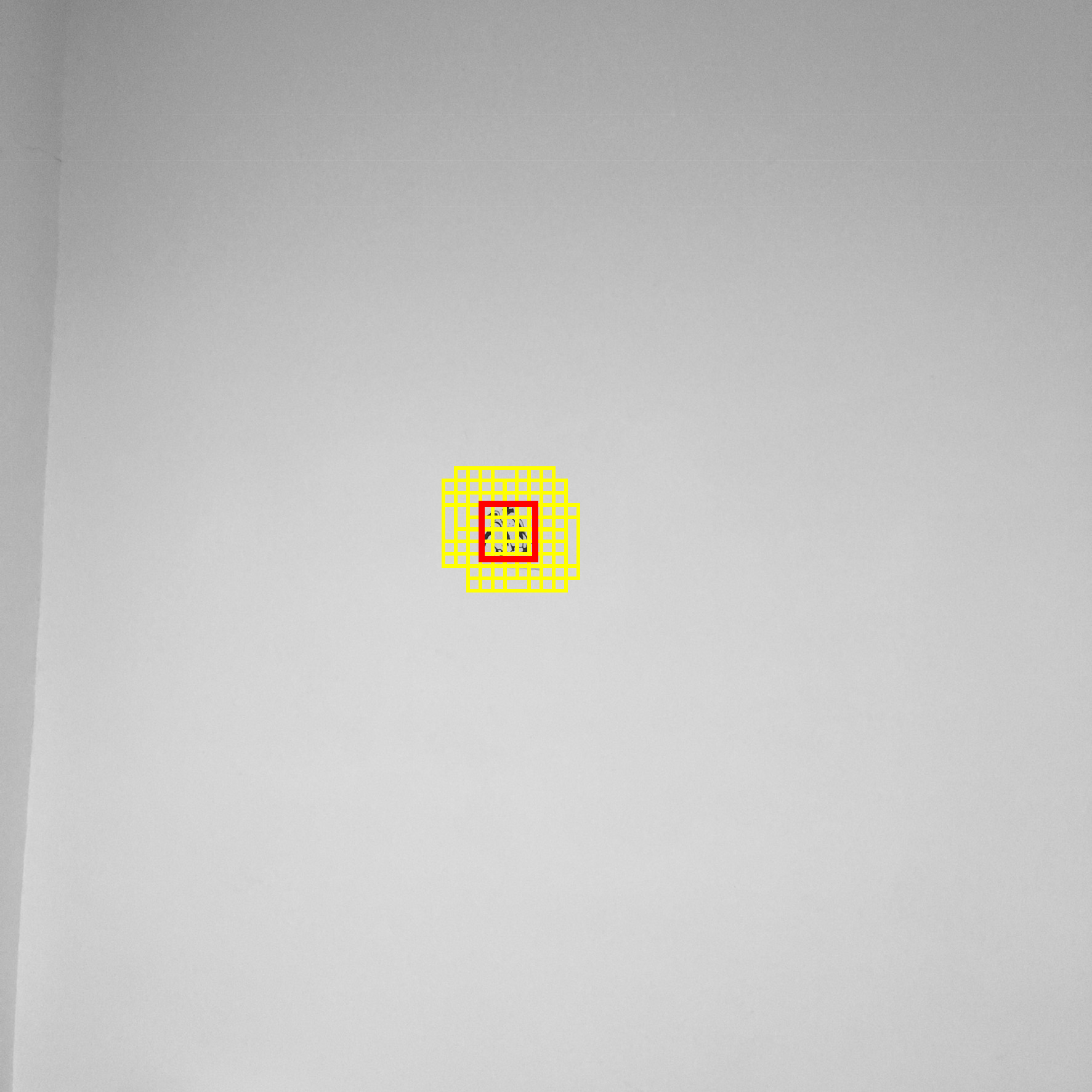}
    }
    \subfloat{
        \includegraphics[width=0.225\linewidth]{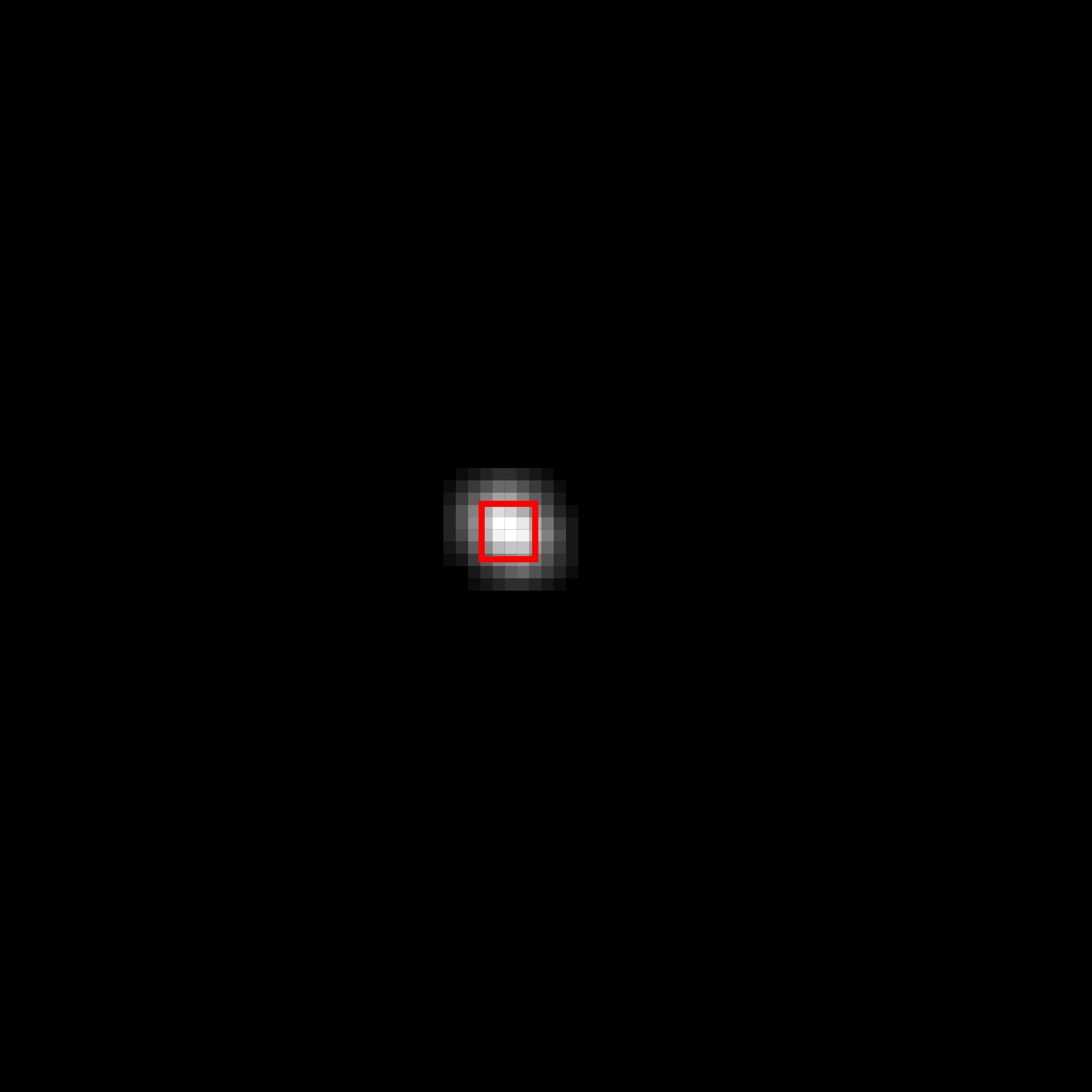}
    }
    \subfloat{
        \includegraphics[width=0.225\linewidth]{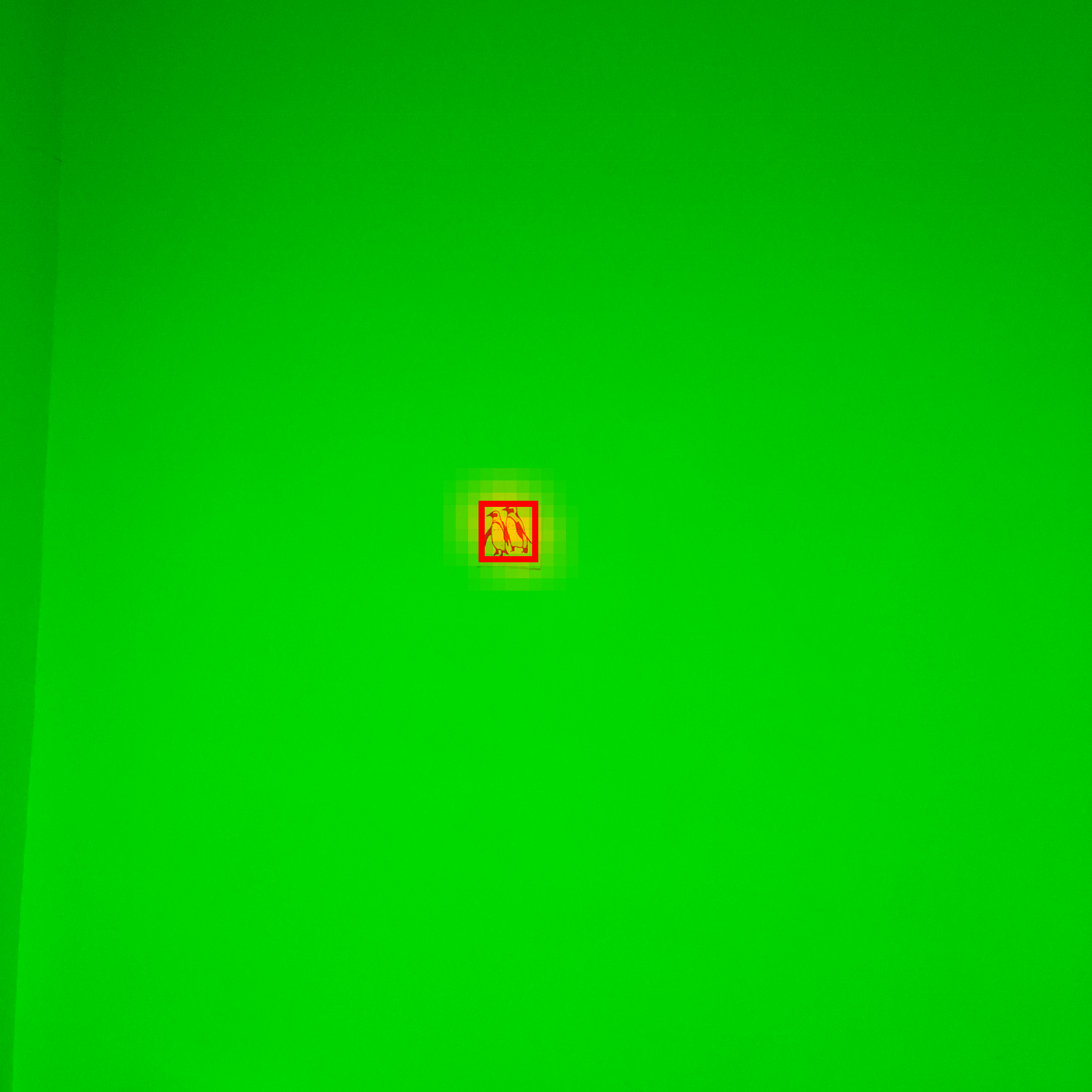}
    }
    \\
    \subfloat{
        \includegraphics[width=0.225\linewidth]{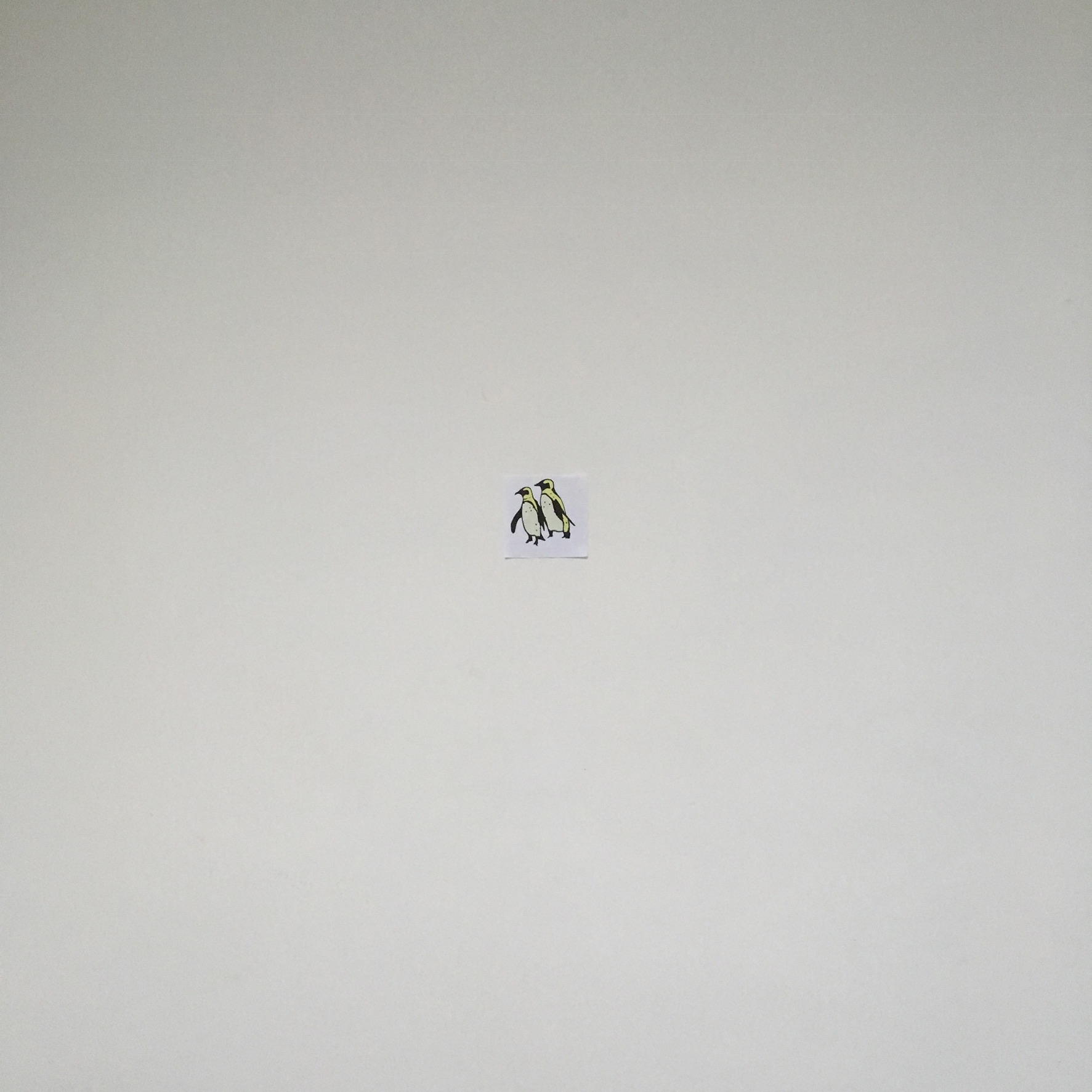}
    }
    \subfloat{
        \includegraphics[width=0.225\linewidth]{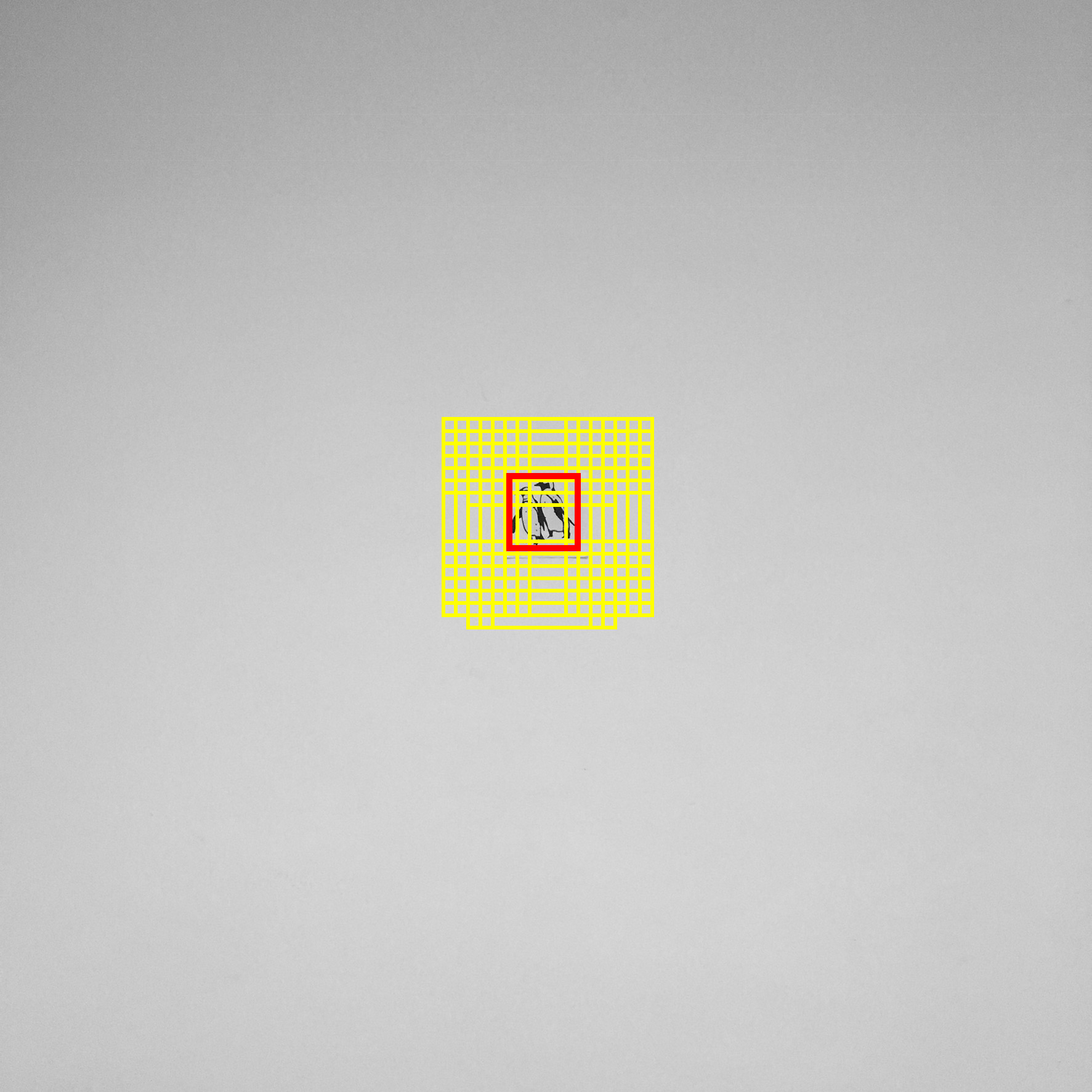}
    }
    \subfloat{
        \includegraphics[width=0.225\linewidth]{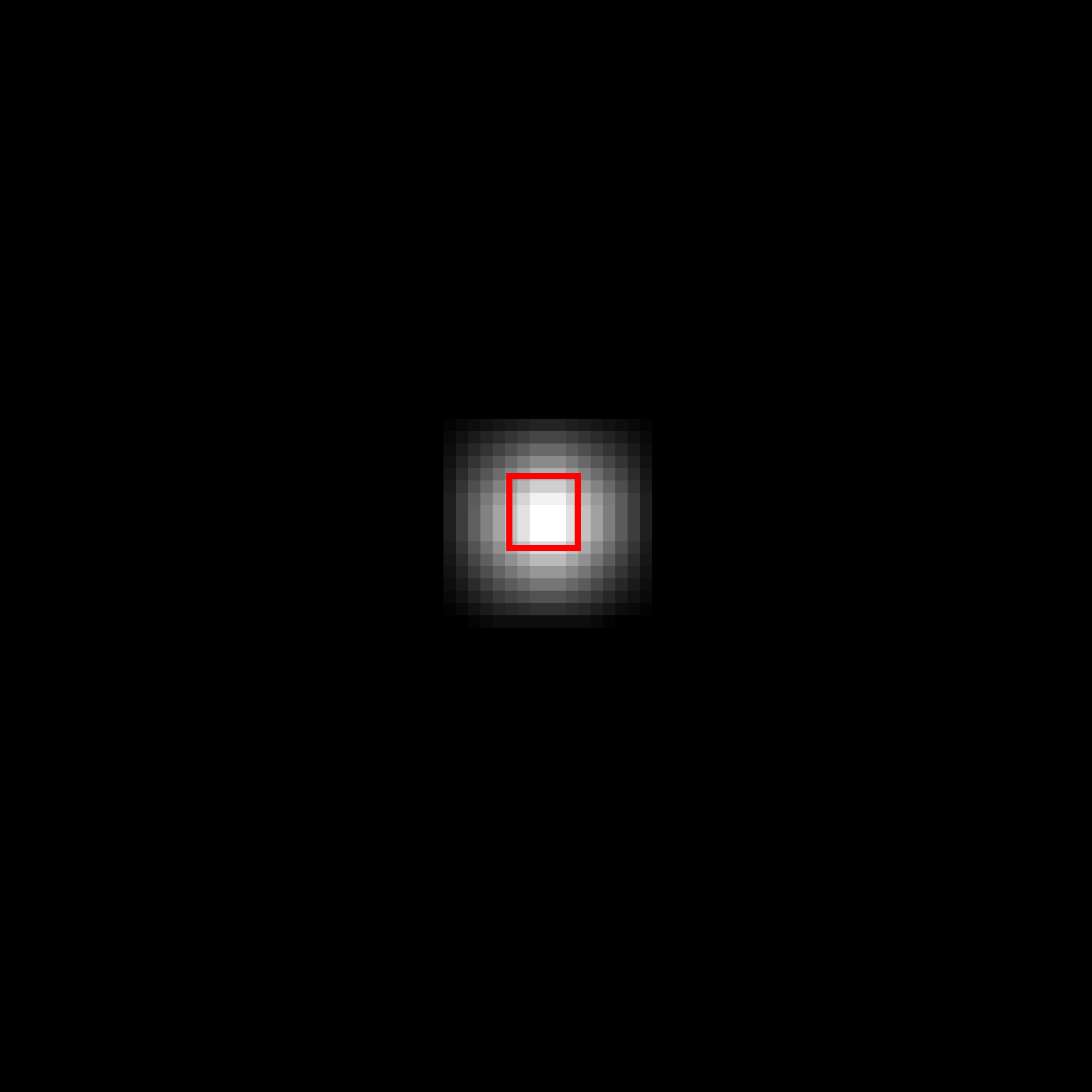}
    }
    \subfloat{
        \includegraphics[width=0.225\linewidth]{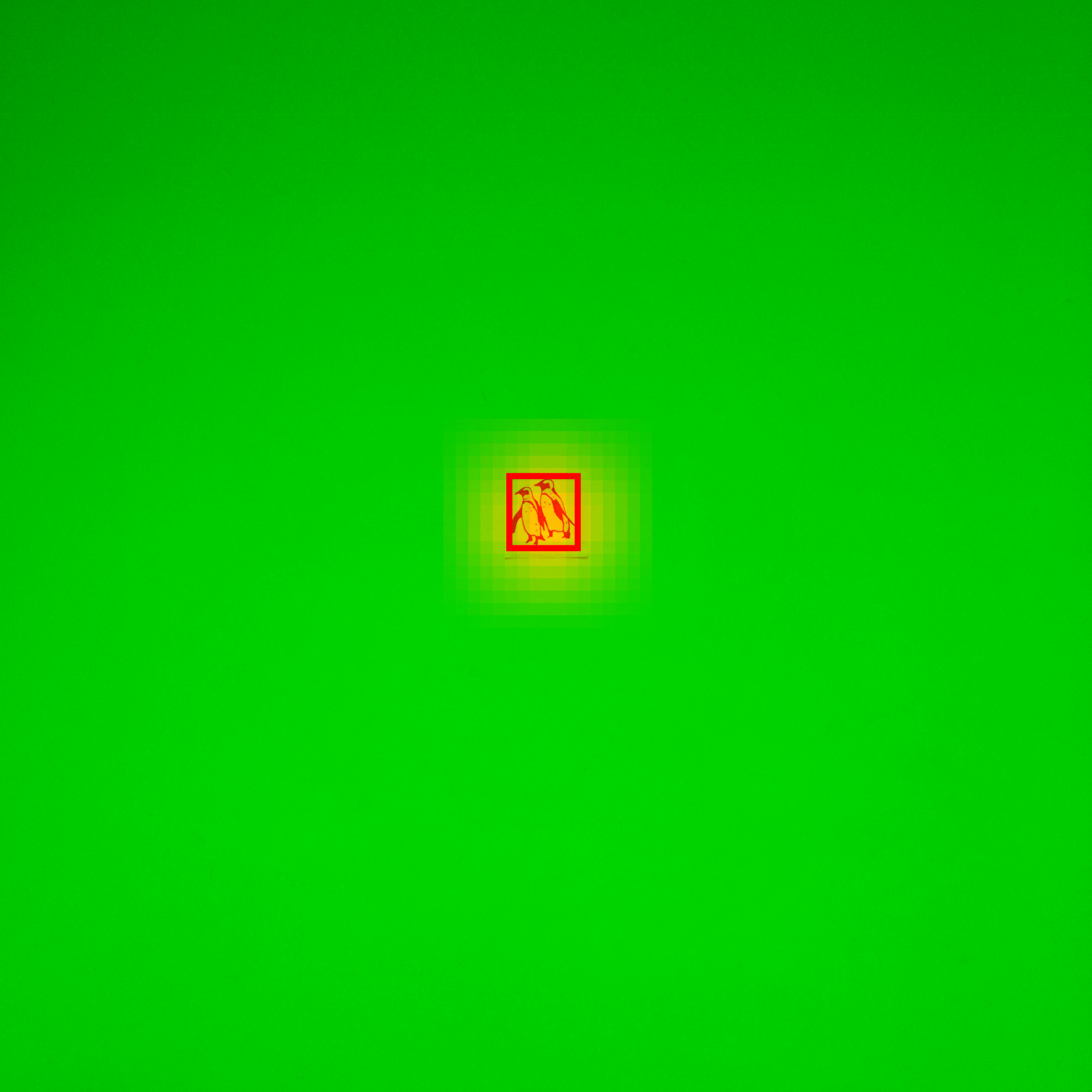}
    }
    \\
    \subfloat{
        \includegraphics[width=0.225\linewidth]{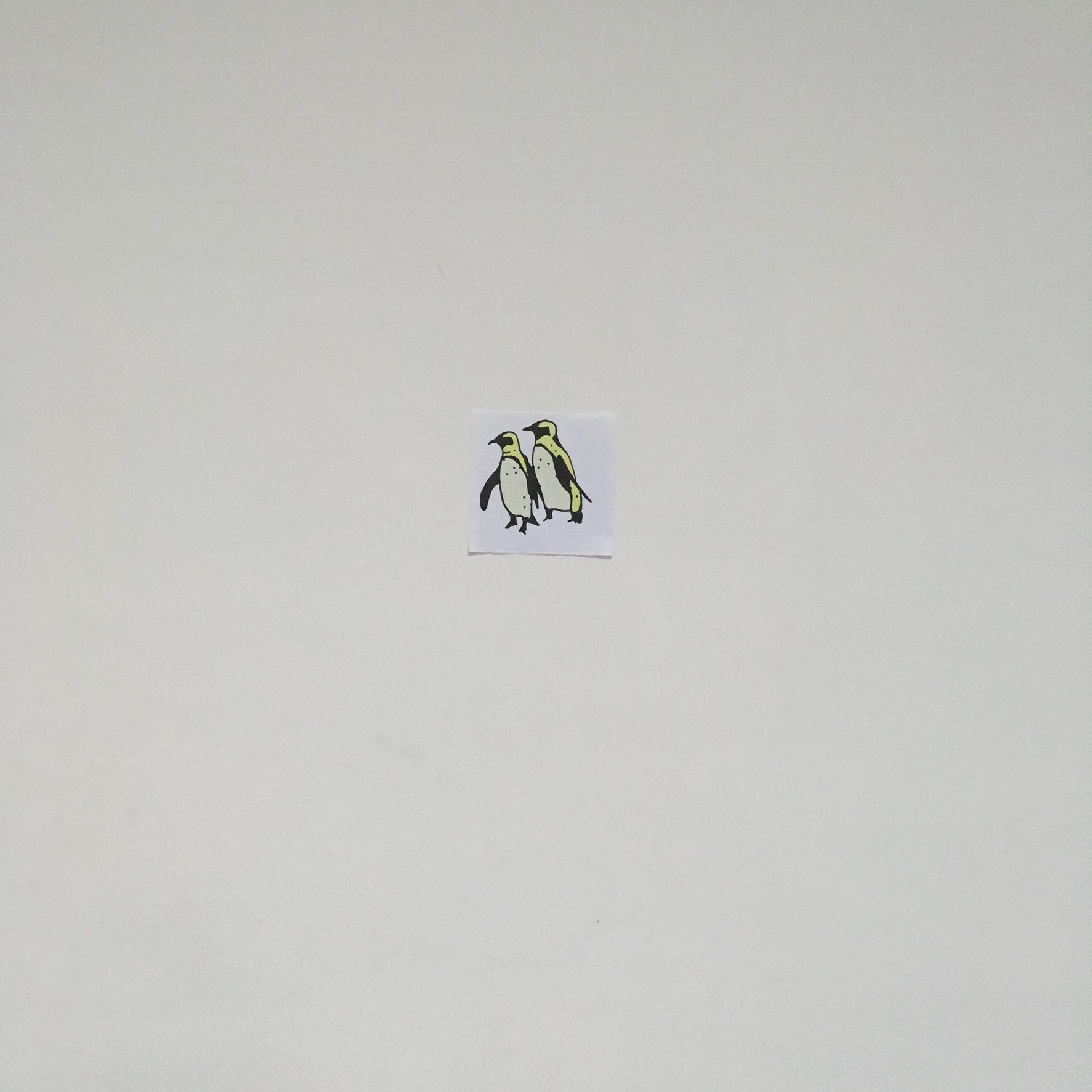}
    }
    \subfloat{
        \includegraphics[width=0.225\linewidth]{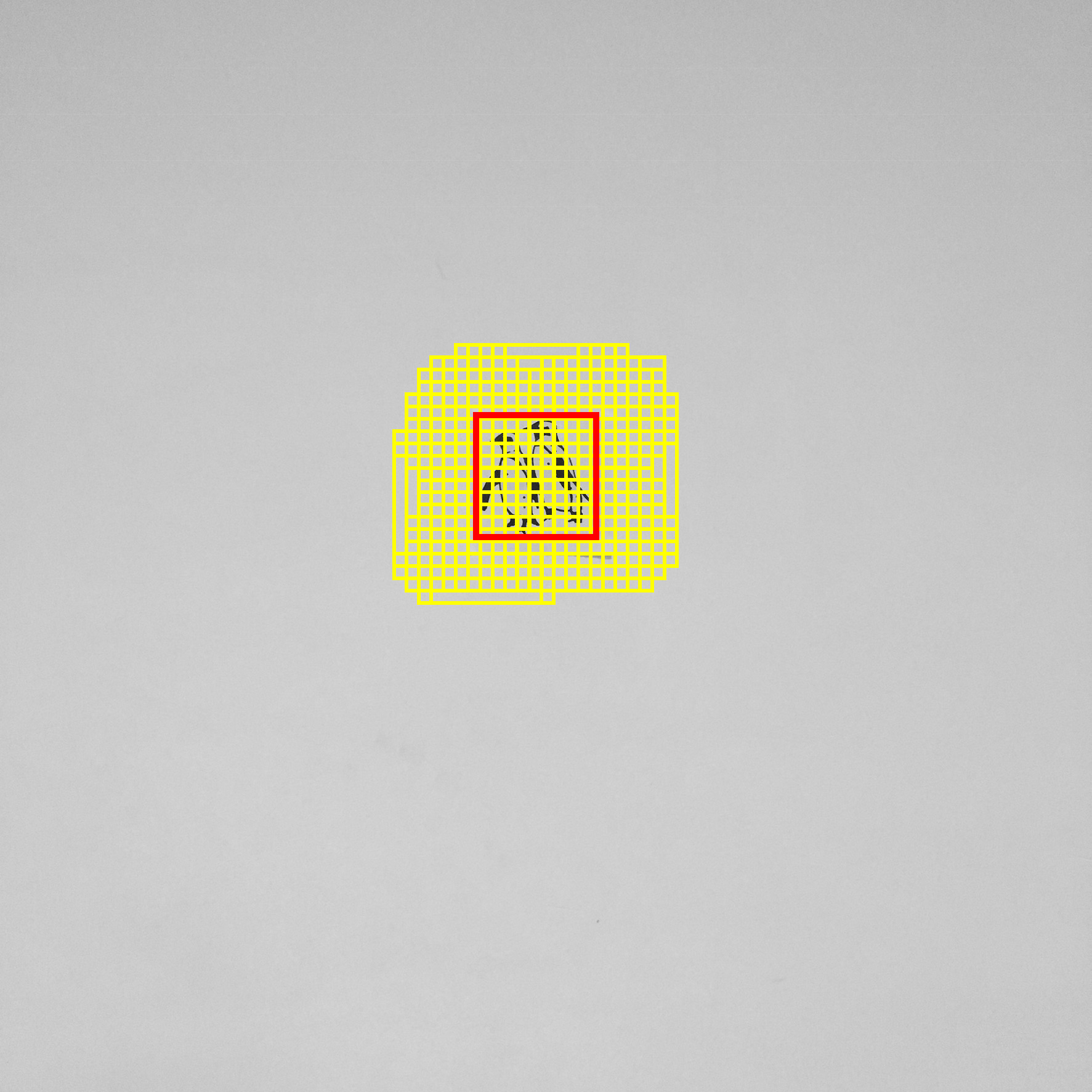}
    }
    \subfloat{
        \includegraphics[width=0.225\linewidth]{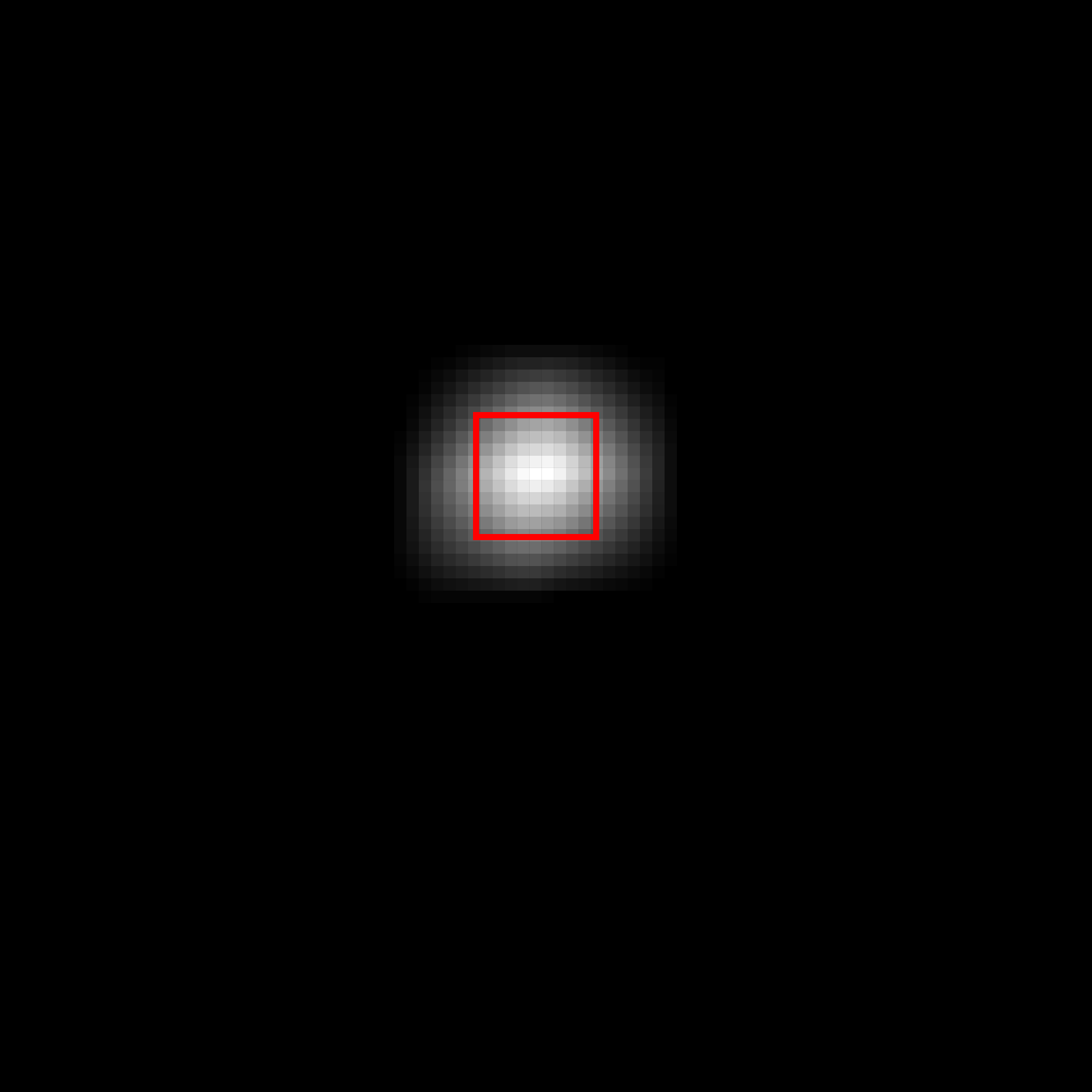}
    }
    \subfloat{
        \includegraphics[width=0.225\linewidth]{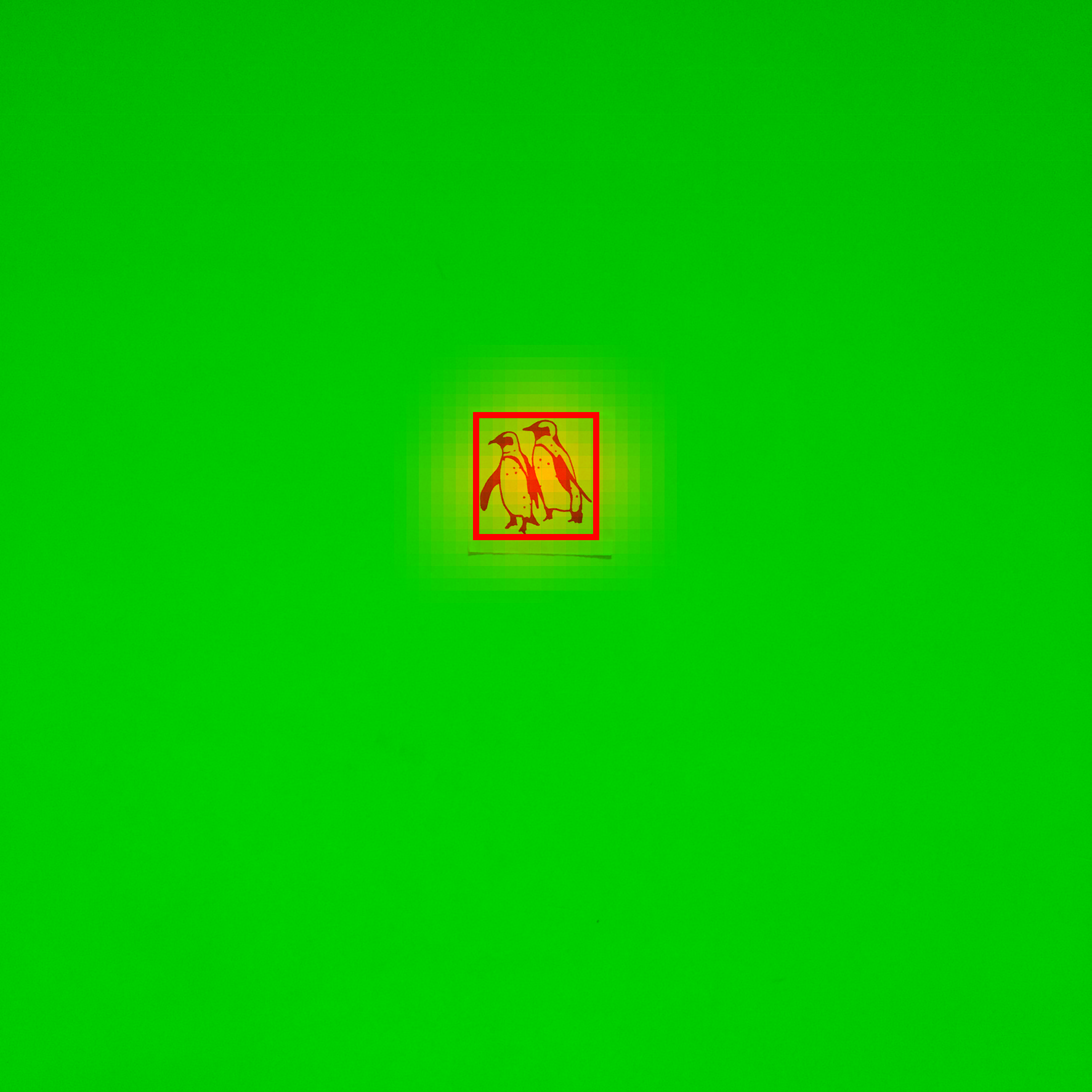}
    }
    \\
    \subfloat{
        \includegraphics[width=0.225\linewidth]{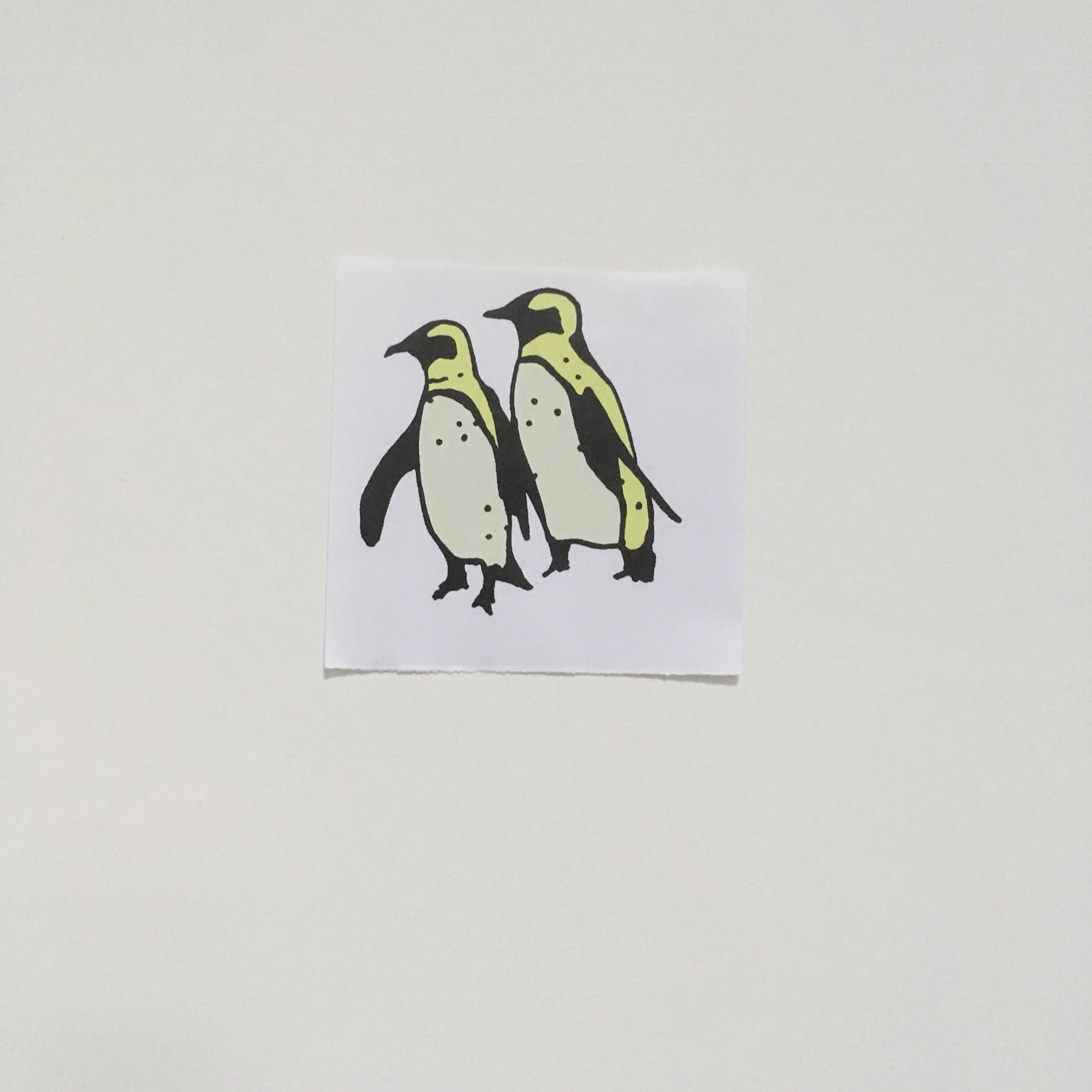}
    }
    \subfloat{
        \includegraphics[width=0.225\linewidth]{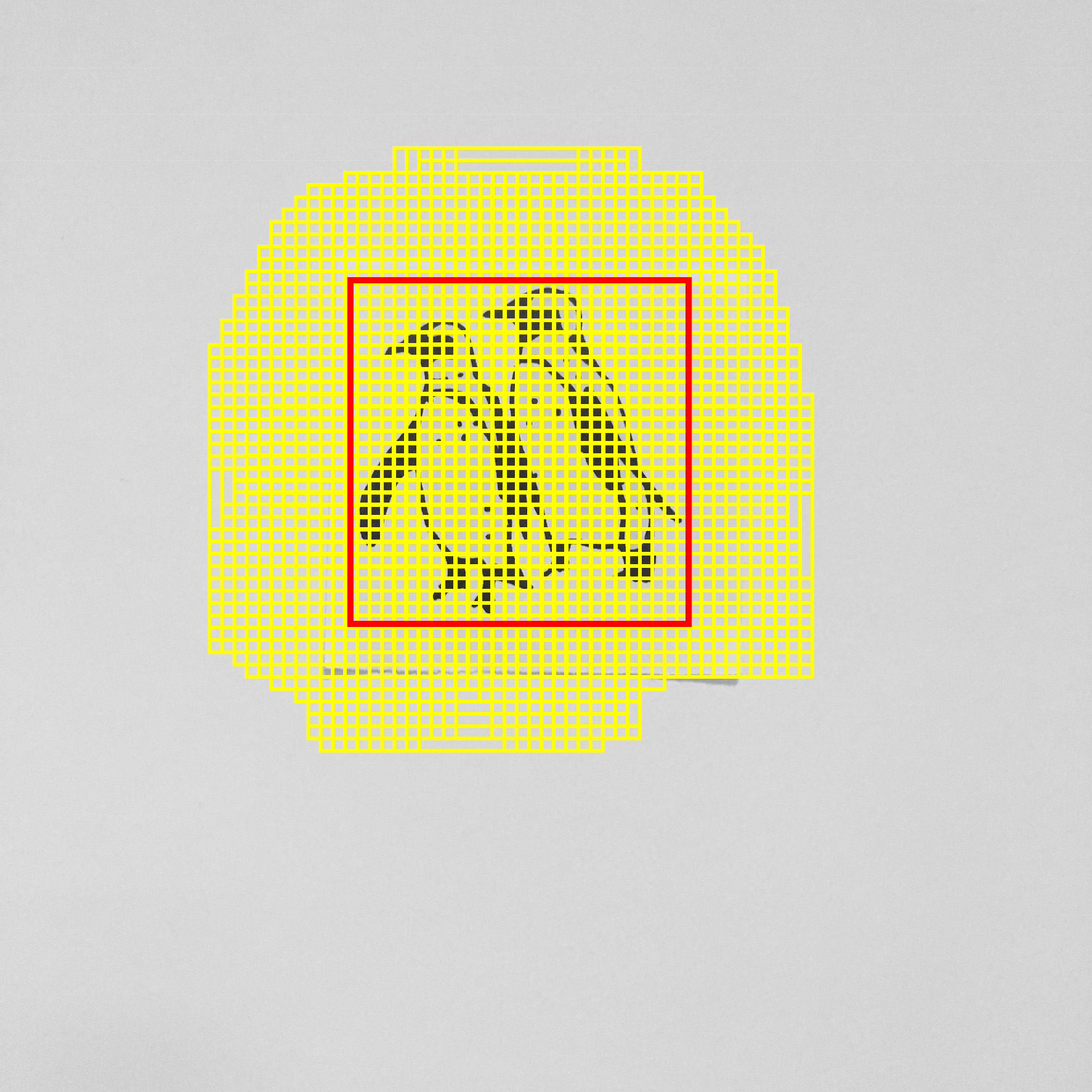}
    }
    \subfloat{
        \includegraphics[width=0.225\linewidth]{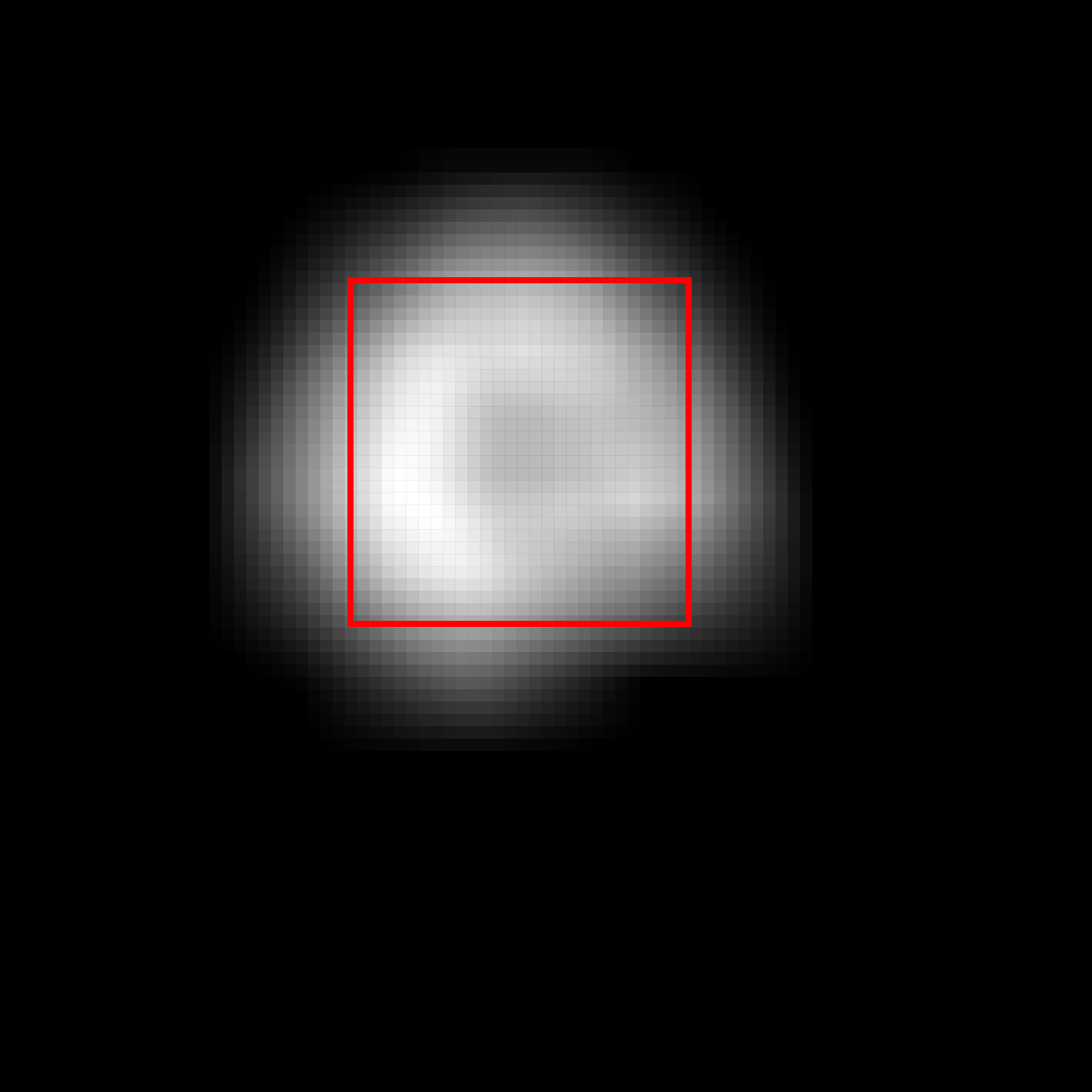}
    }
    \subfloat{
        \includegraphics[width=0.225\linewidth]{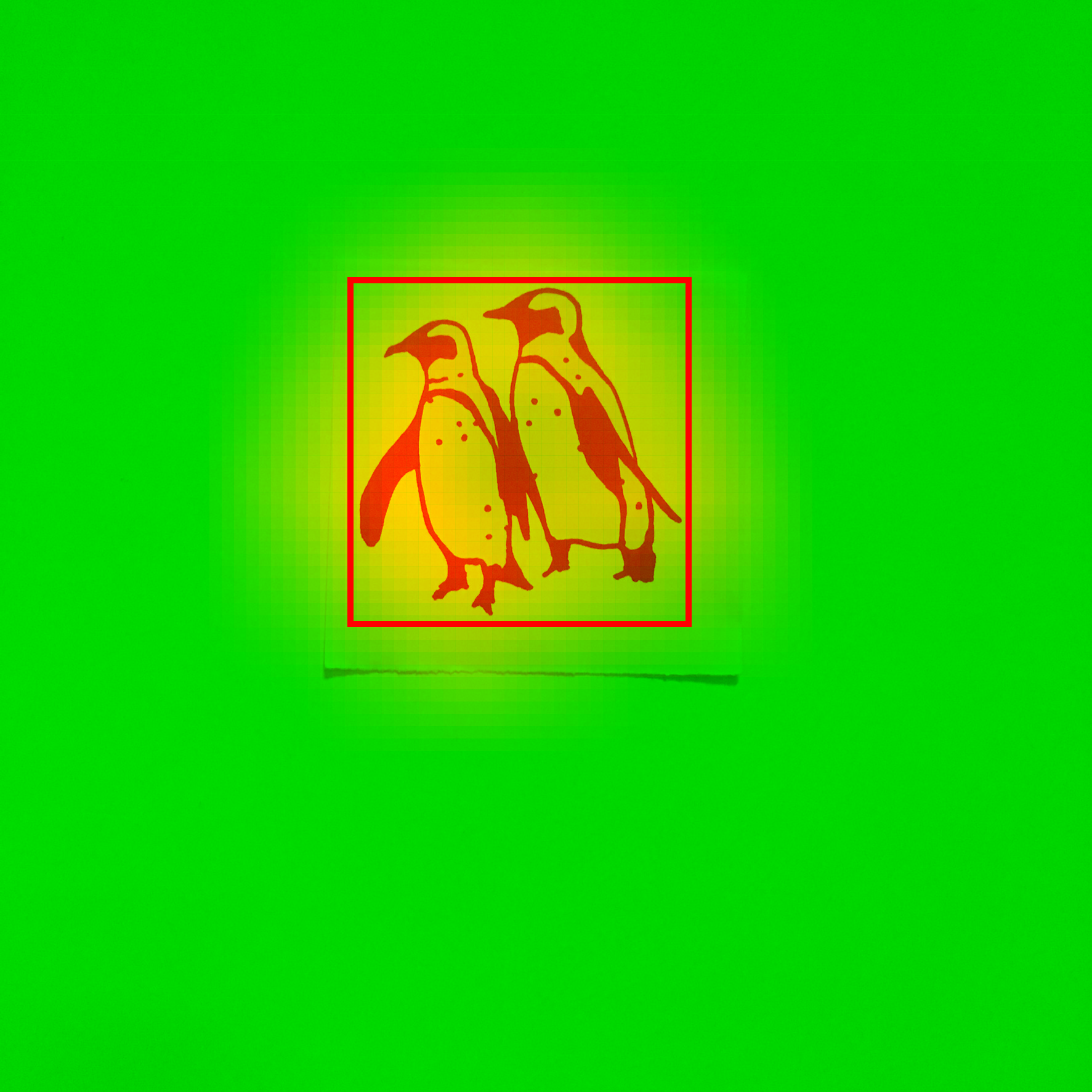}
    }
    \\
    \subfloat[Input\label{subfig:study1_input}]{
        \includegraphics[width=0.225\linewidth]{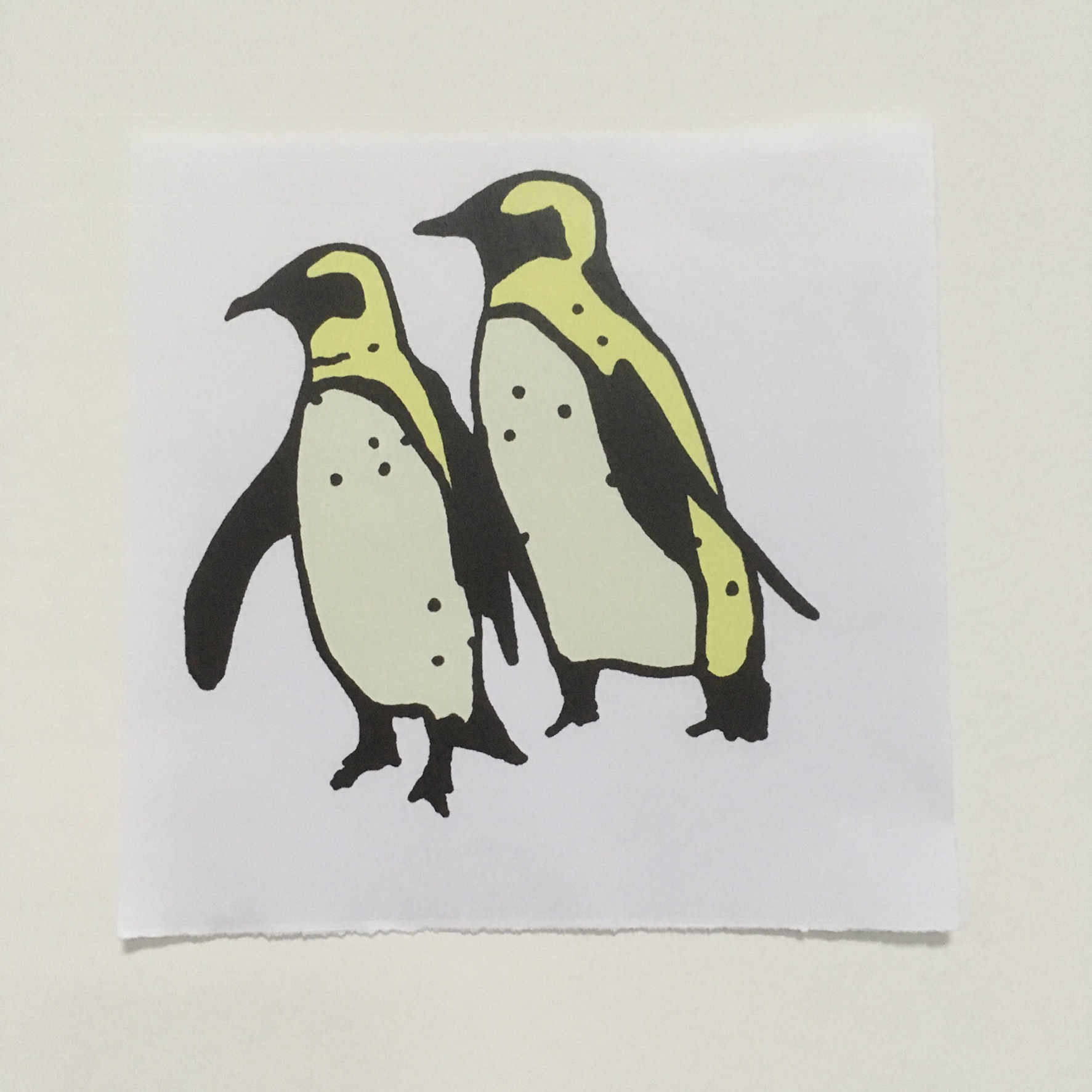}
    }
    \subfloat[Proposal\label{subfig:study1_proposal}]{
        \includegraphics[width=0.225\linewidth]{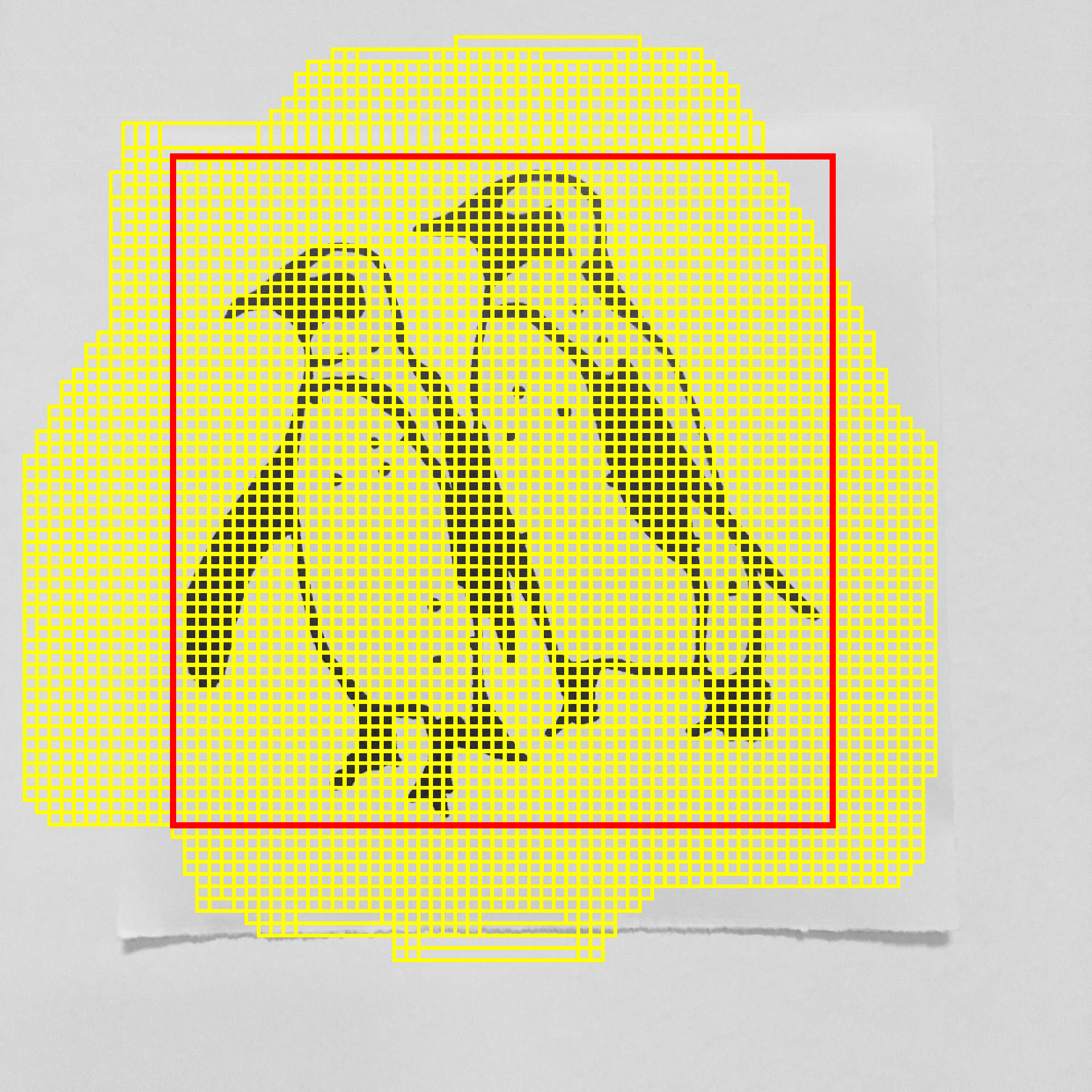}
    }
    \subfloat[Gray\label{subfig:study1_gray}]{
        \includegraphics[width=0.225\linewidth]{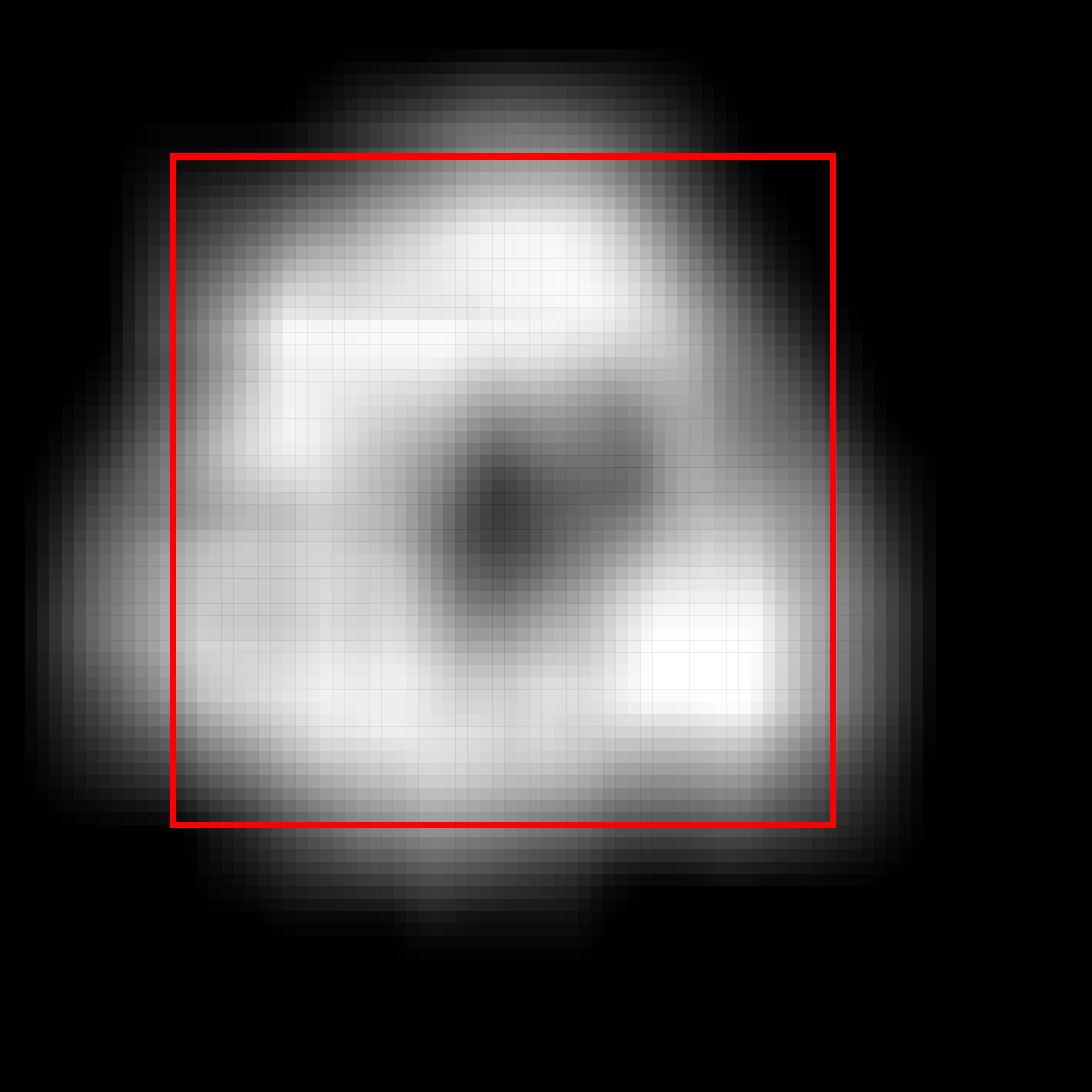}
    }
    \subfloat[Fused\label{subfig:study1_fused}]{
        \includegraphics[width=0.225\linewidth]{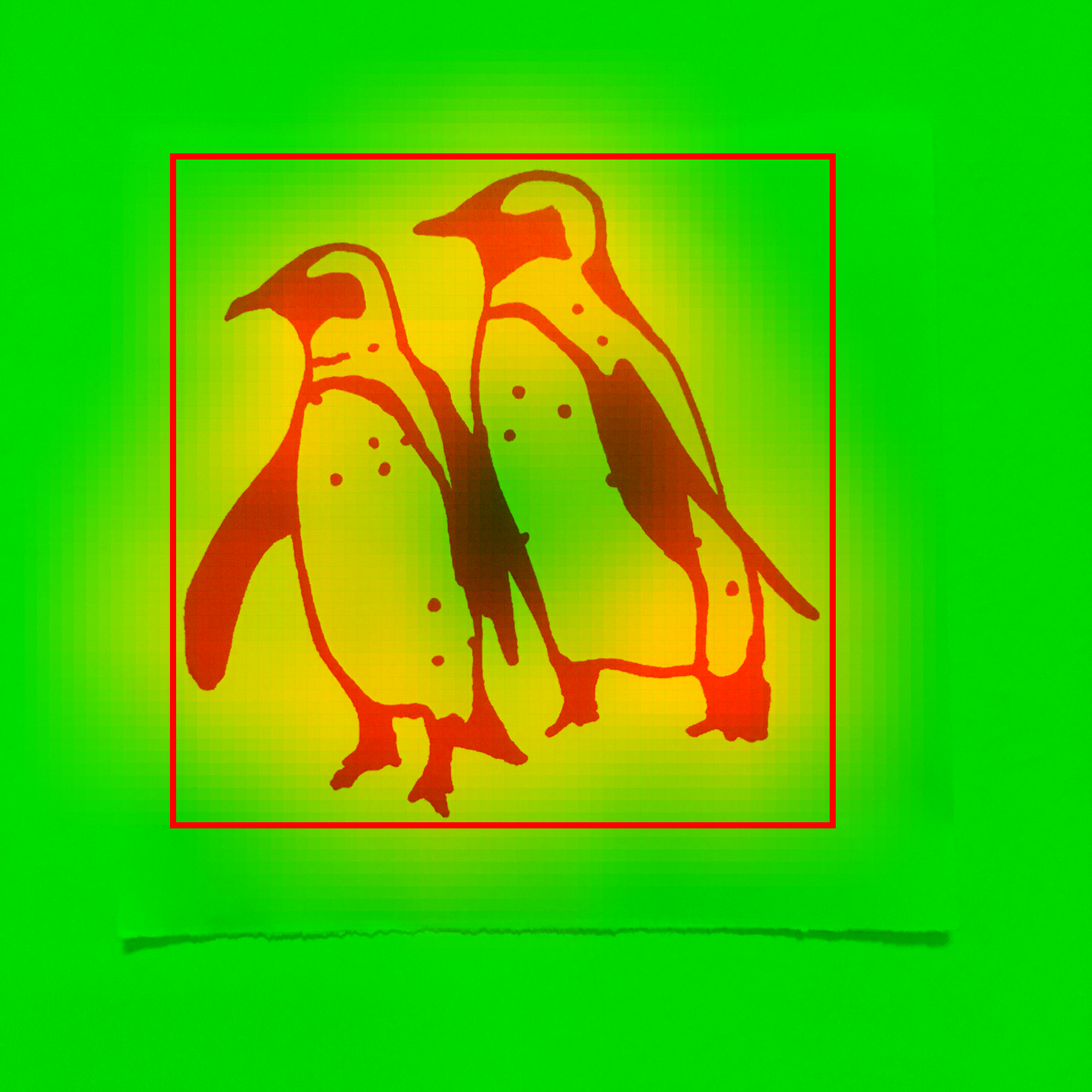}
    }
    \caption{Simple Artcode detection study in clean background, good lighting.}
    \label{fig:study_1} 
\end{figure}

\begin{figure} [t]
    \centering
    \addtocounter{subfigure}{-16}
    \subfloat{
        \includegraphics[width=0.225\linewidth]{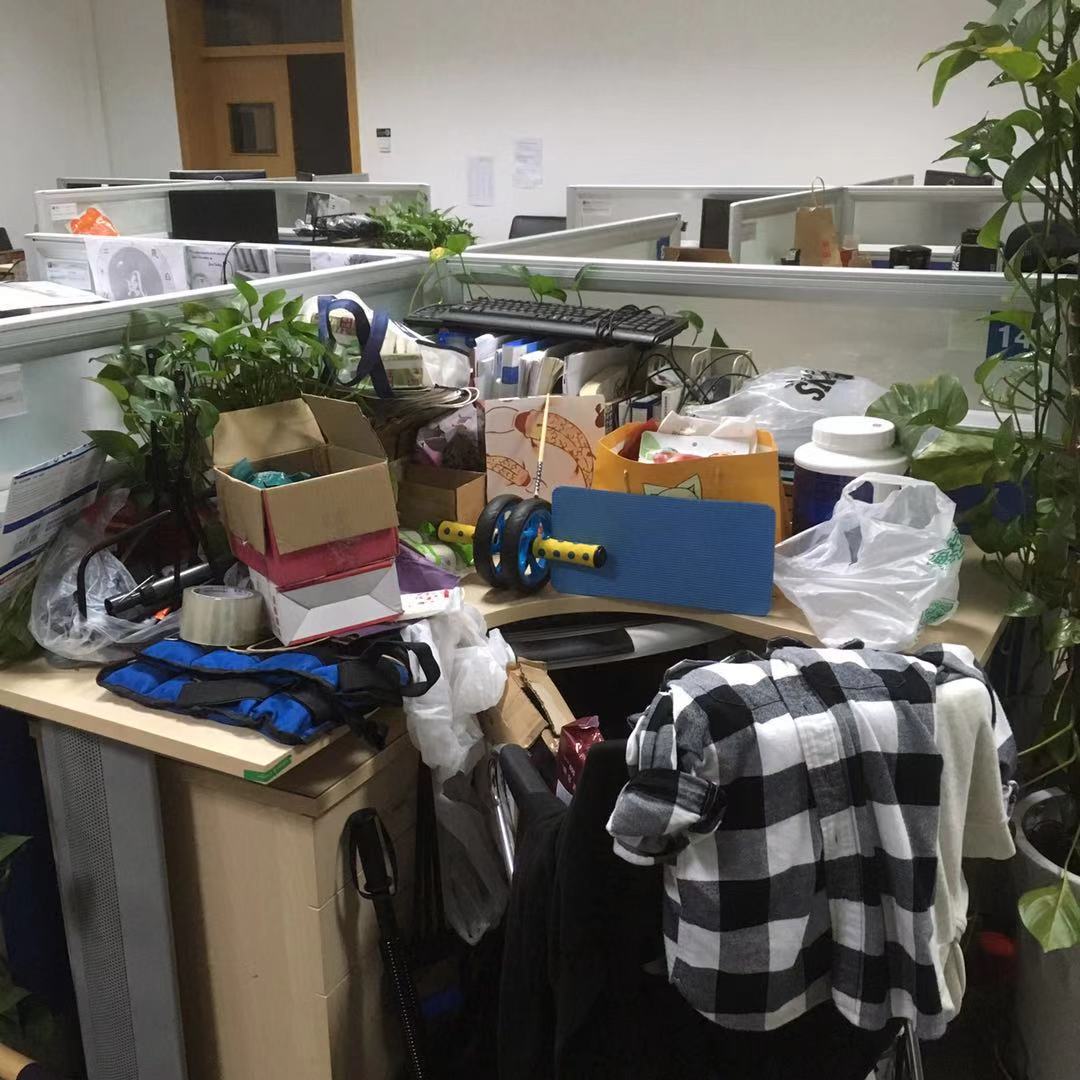}
    }
    \subfloat{
        \includegraphics[width=0.225\linewidth]{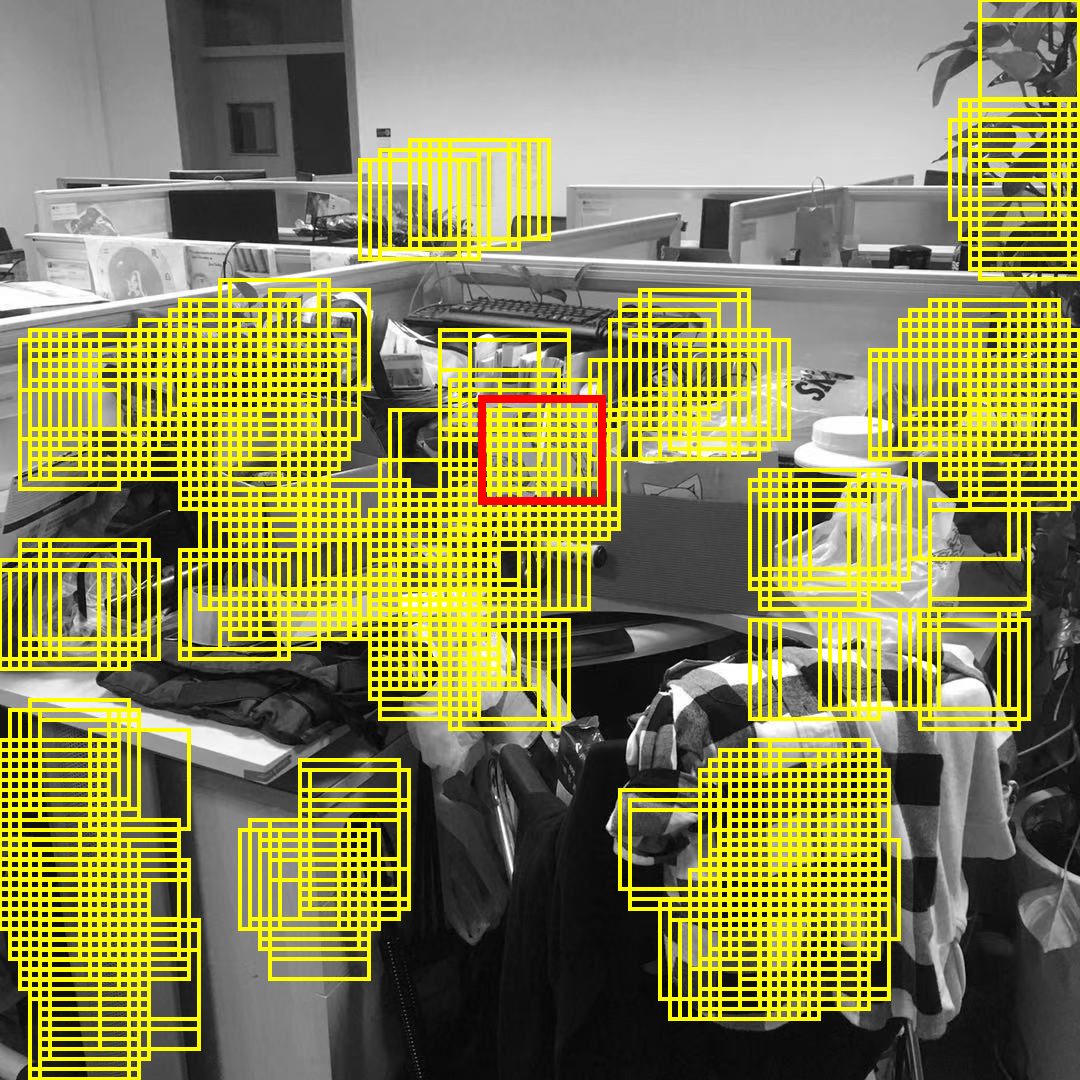}
    }
    \subfloat{
        \includegraphics[width=0.225\linewidth]{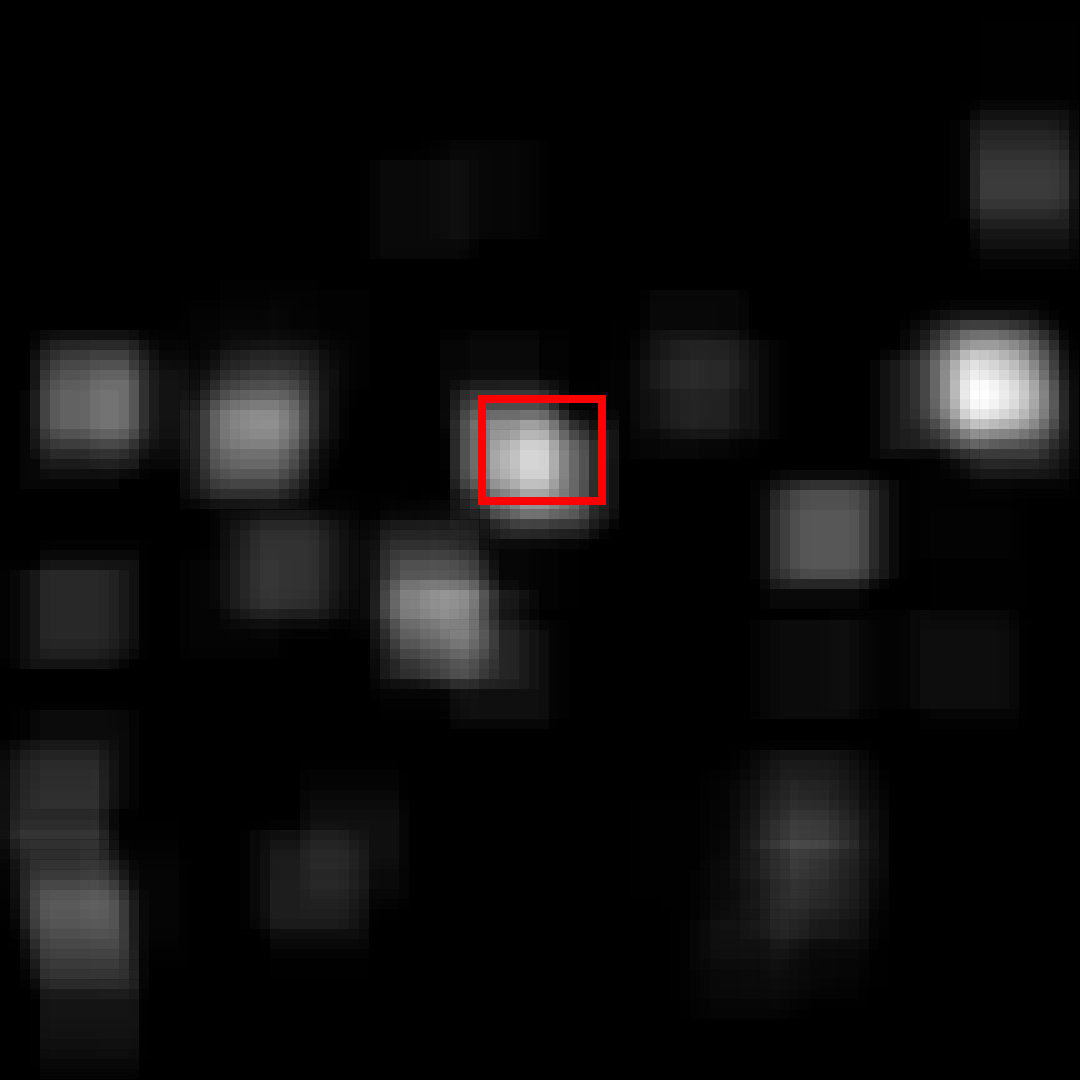}
    }
    \subfloat{
        \includegraphics[width=0.225\linewidth]{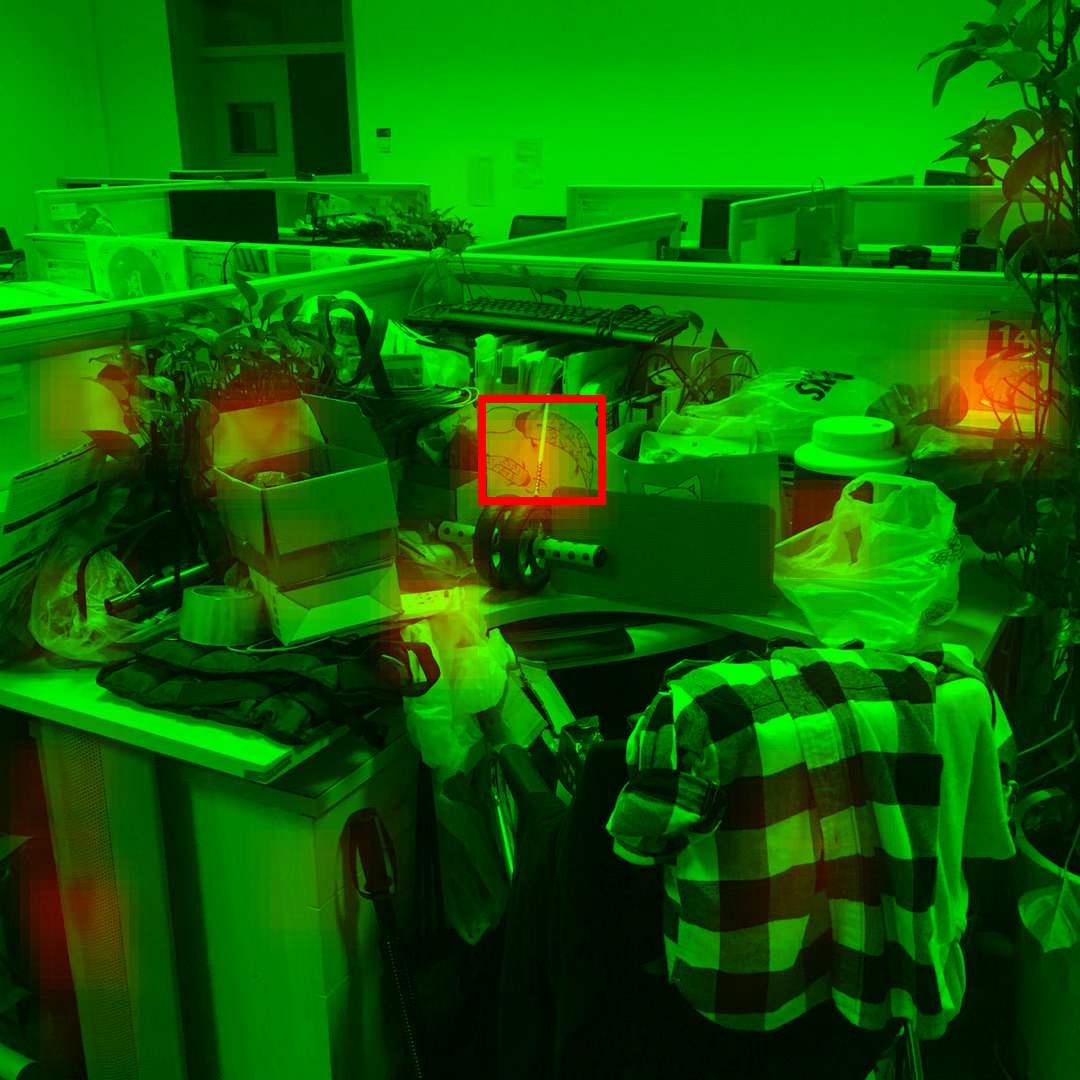}
    }
    \\
    \subfloat{
        \includegraphics[width=0.225\linewidth]{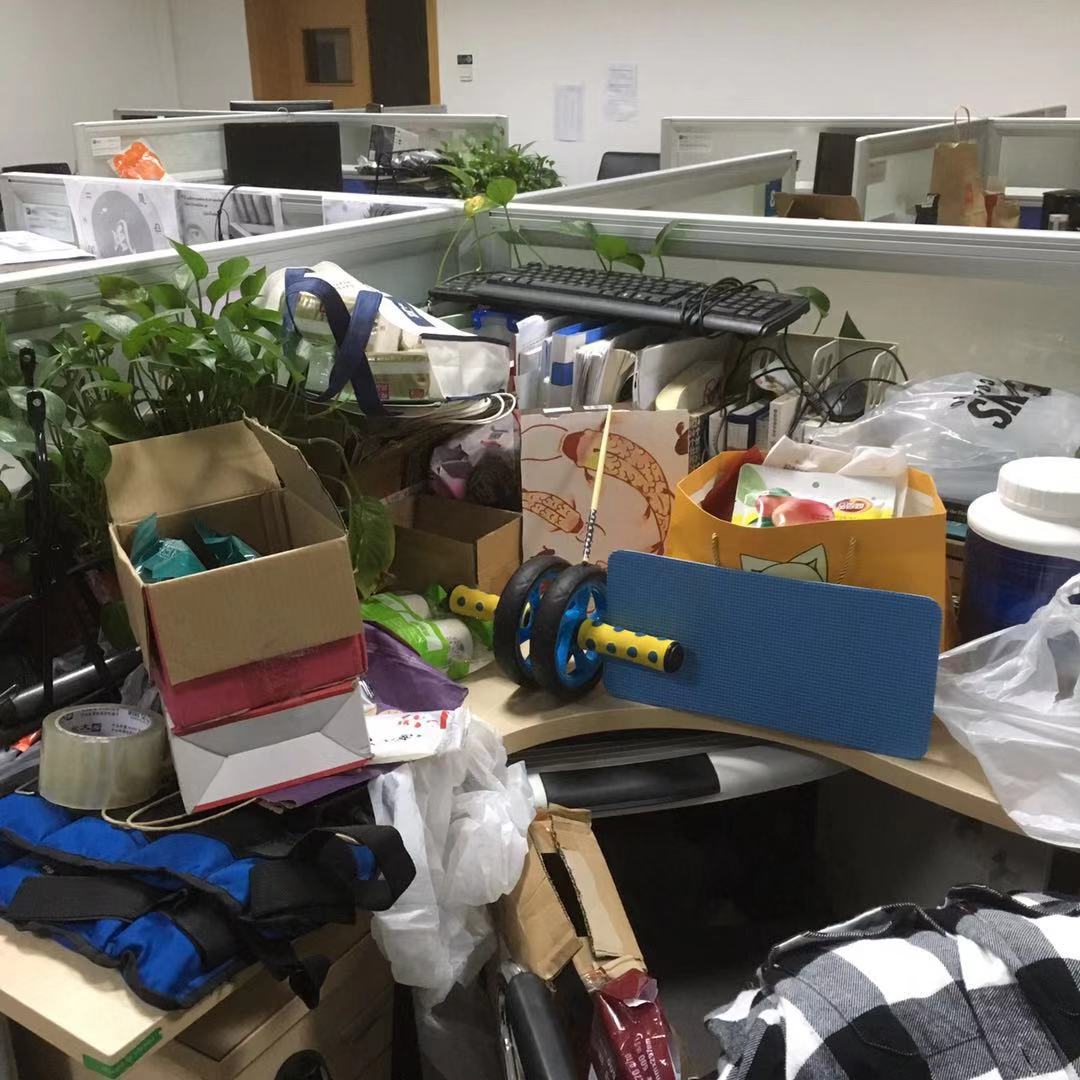}
    }
    \subfloat{
        \includegraphics[width=0.225\linewidth]{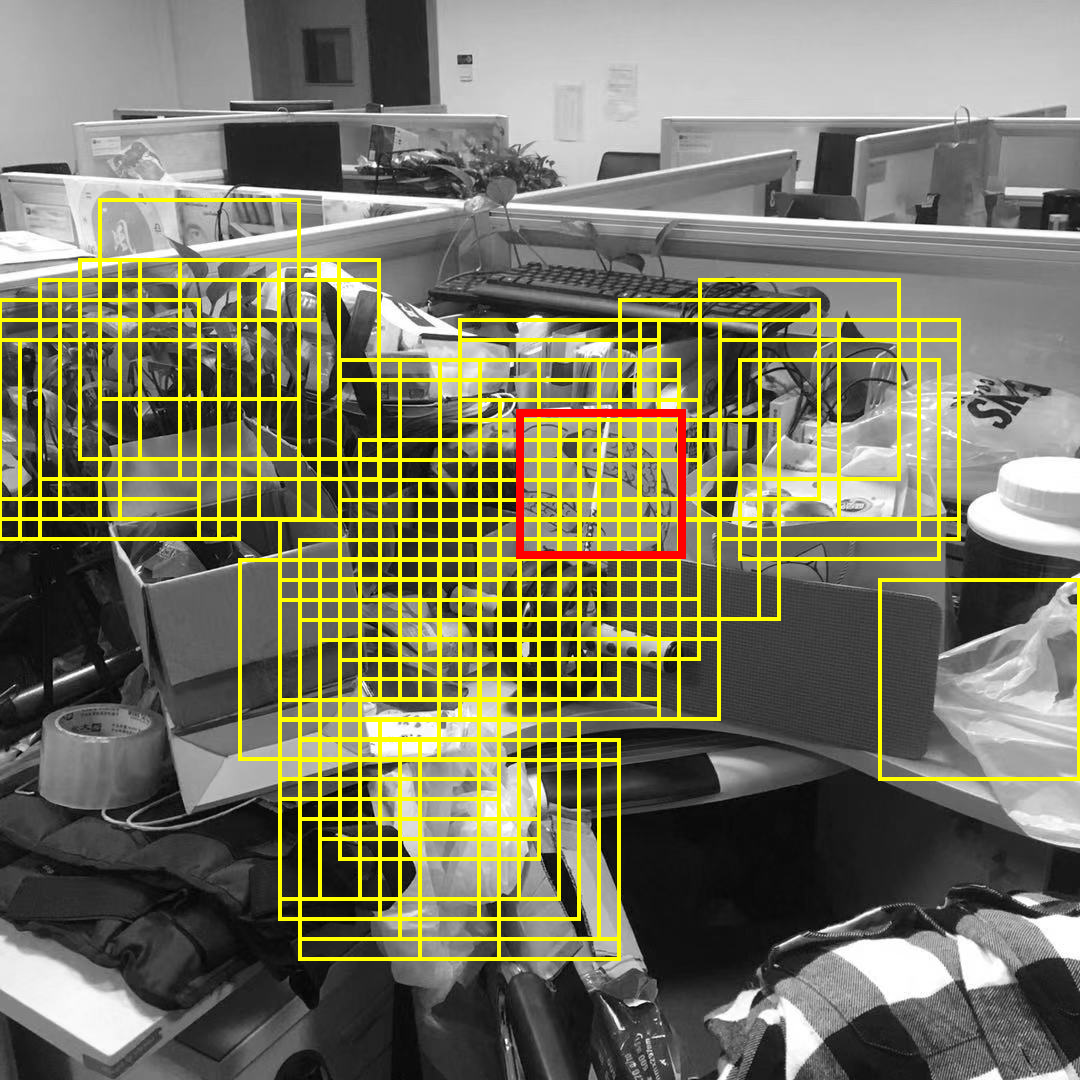}
    }
    \subfloat{
        \includegraphics[width=0.225\linewidth]{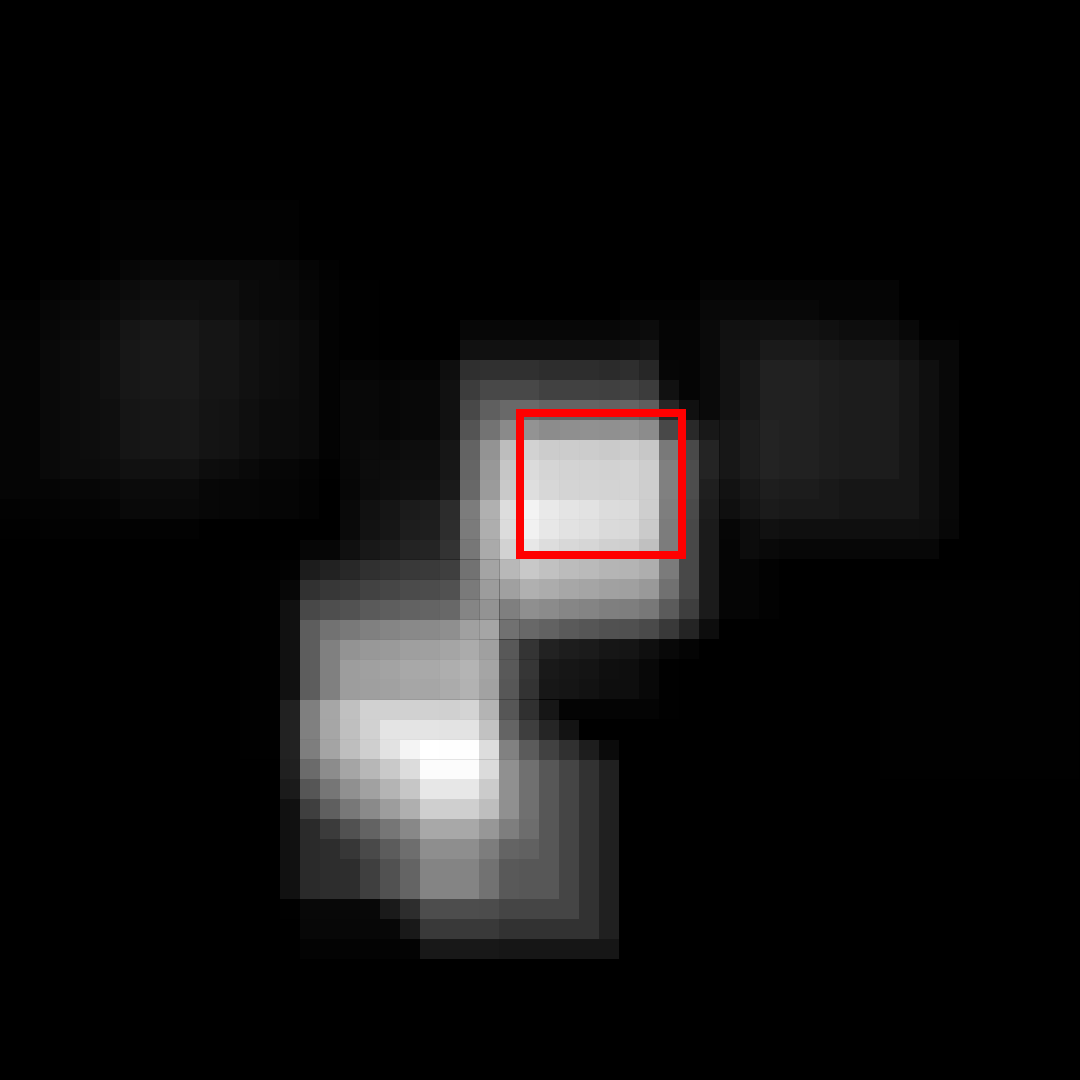}
    }
    \subfloat{
        \includegraphics[width=0.225\linewidth]{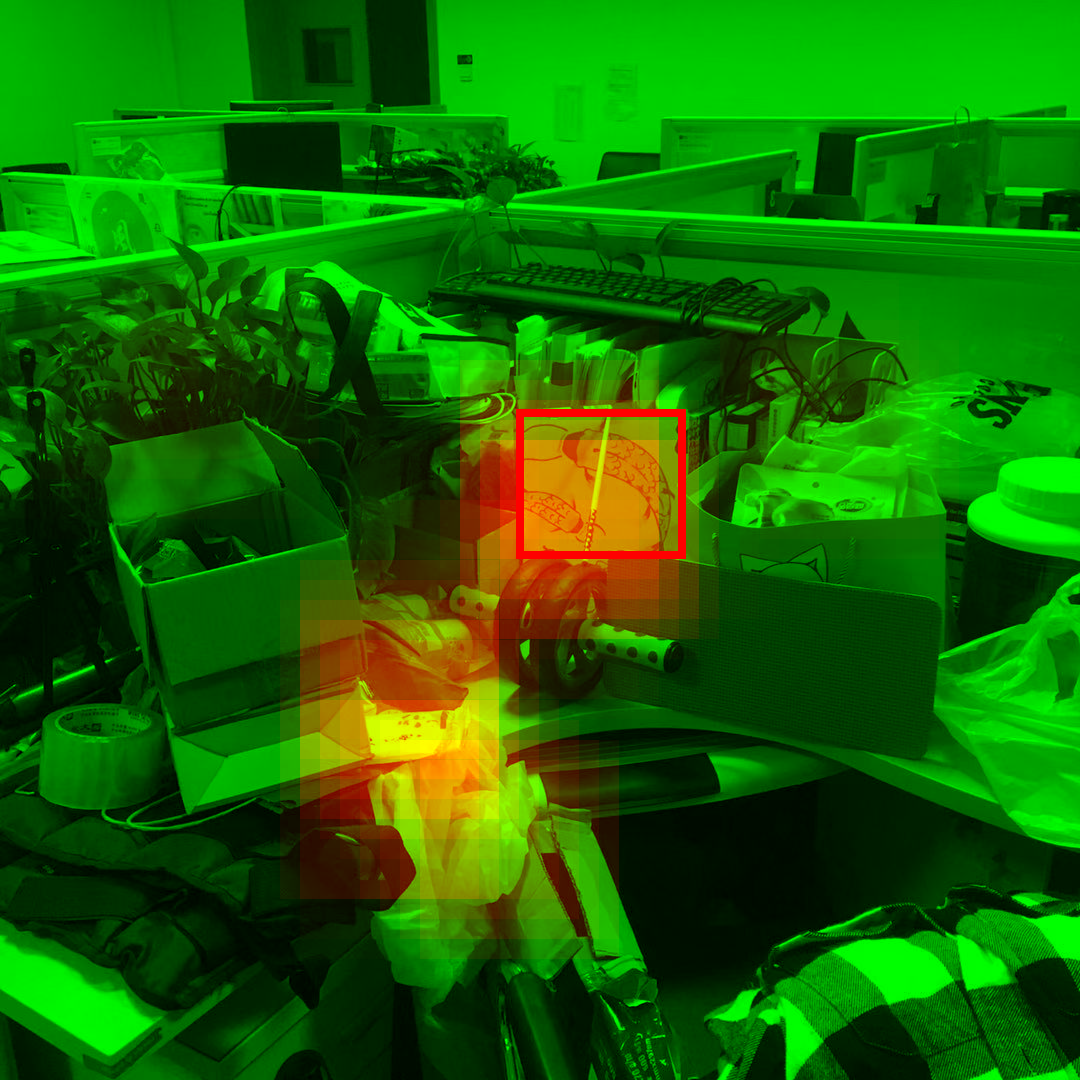}
    }
    \\
    \subfloat{
        \includegraphics[width=0.225\linewidth]{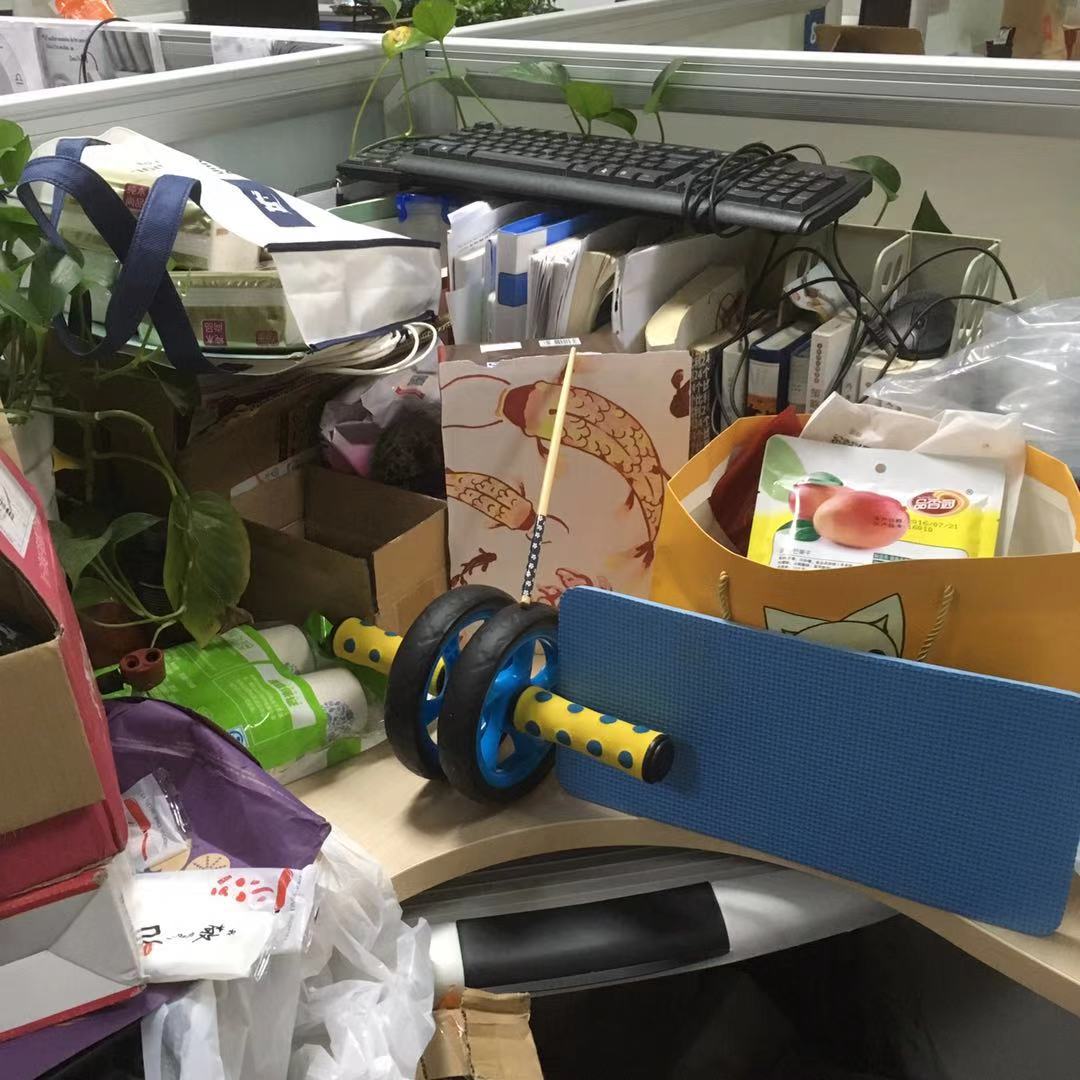}
    }
    \subfloat{
        \includegraphics[width=0.225\linewidth]{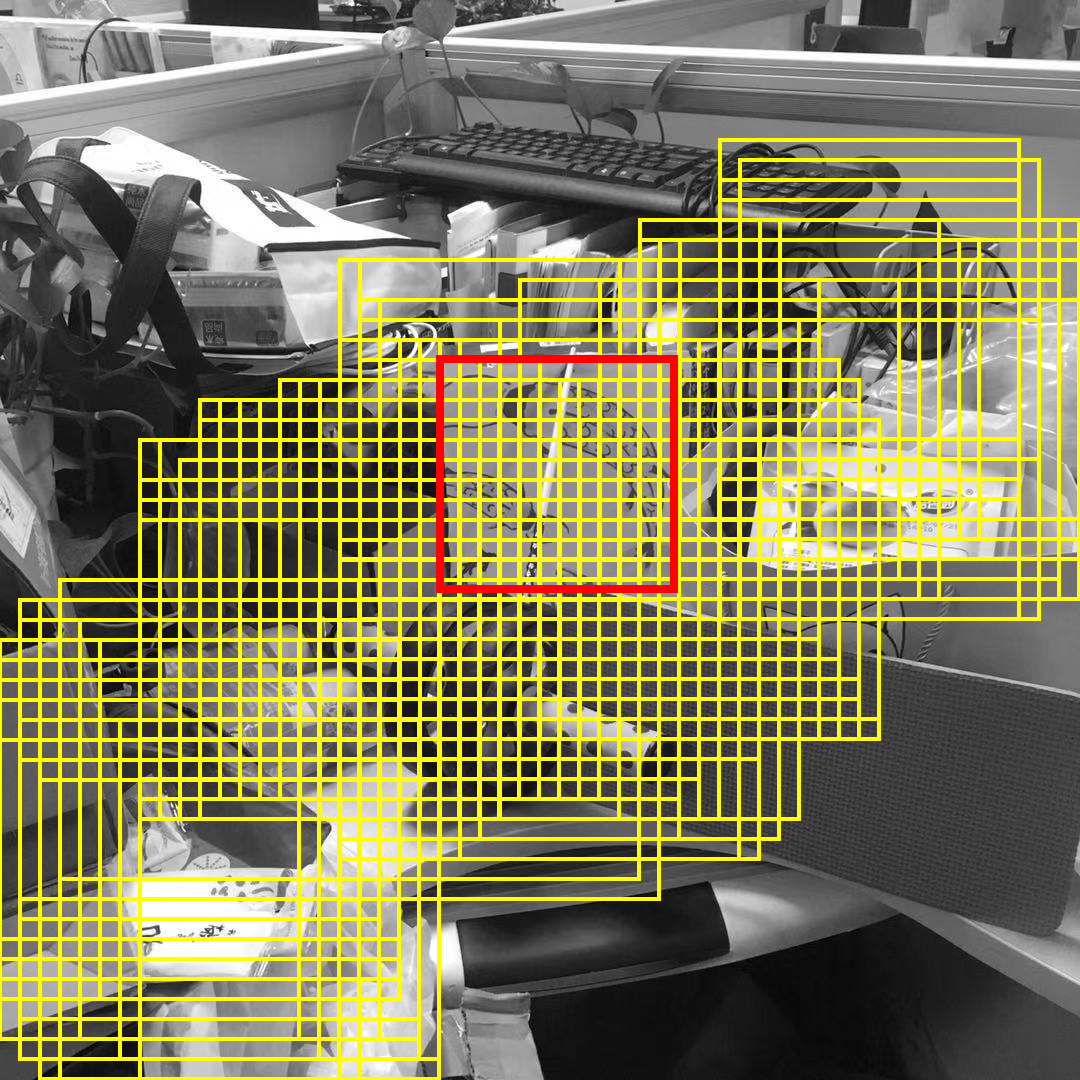}
    }
    \subfloat{
        \includegraphics[width=0.225\linewidth]{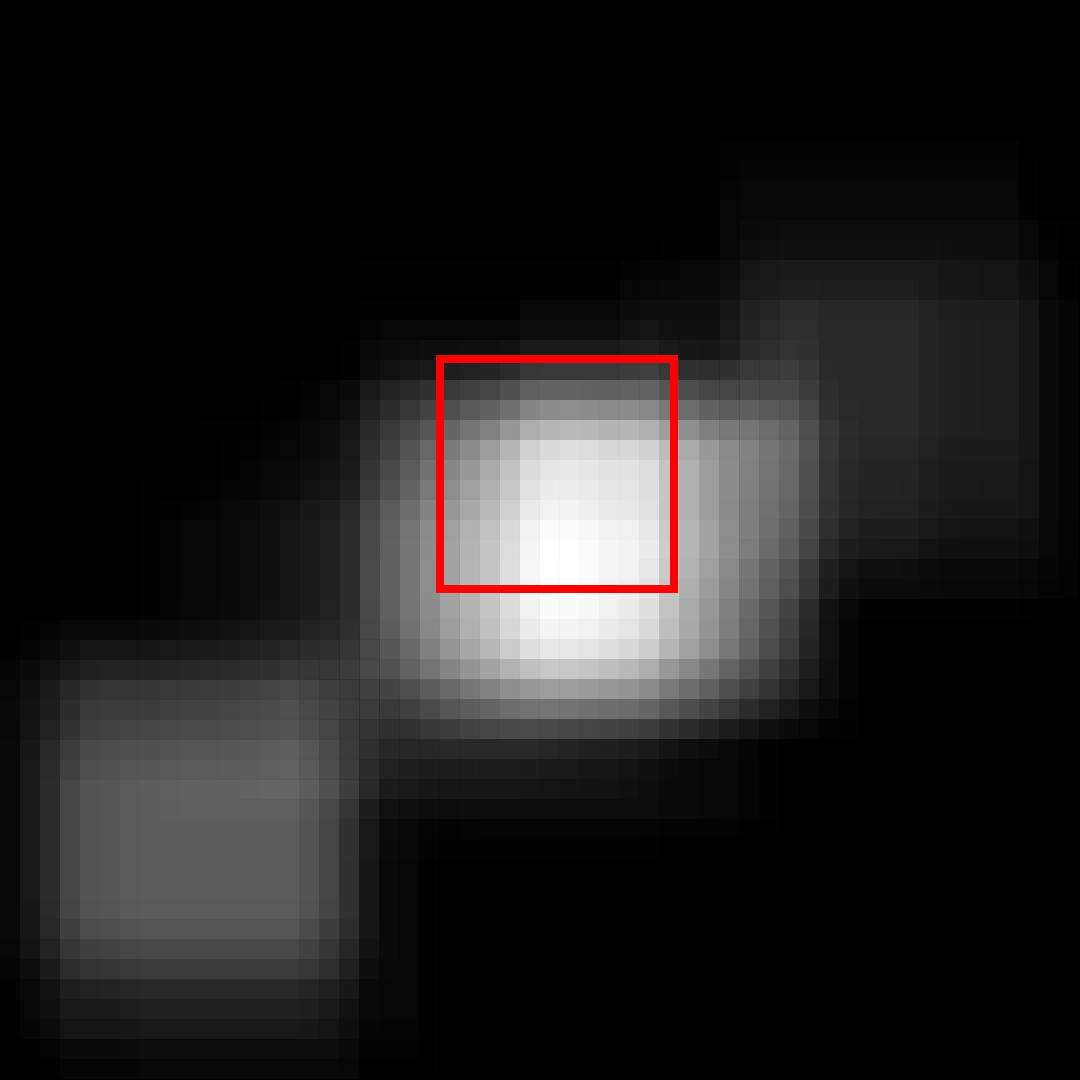}
    }
    \subfloat{
        \includegraphics[width=0.225\linewidth]{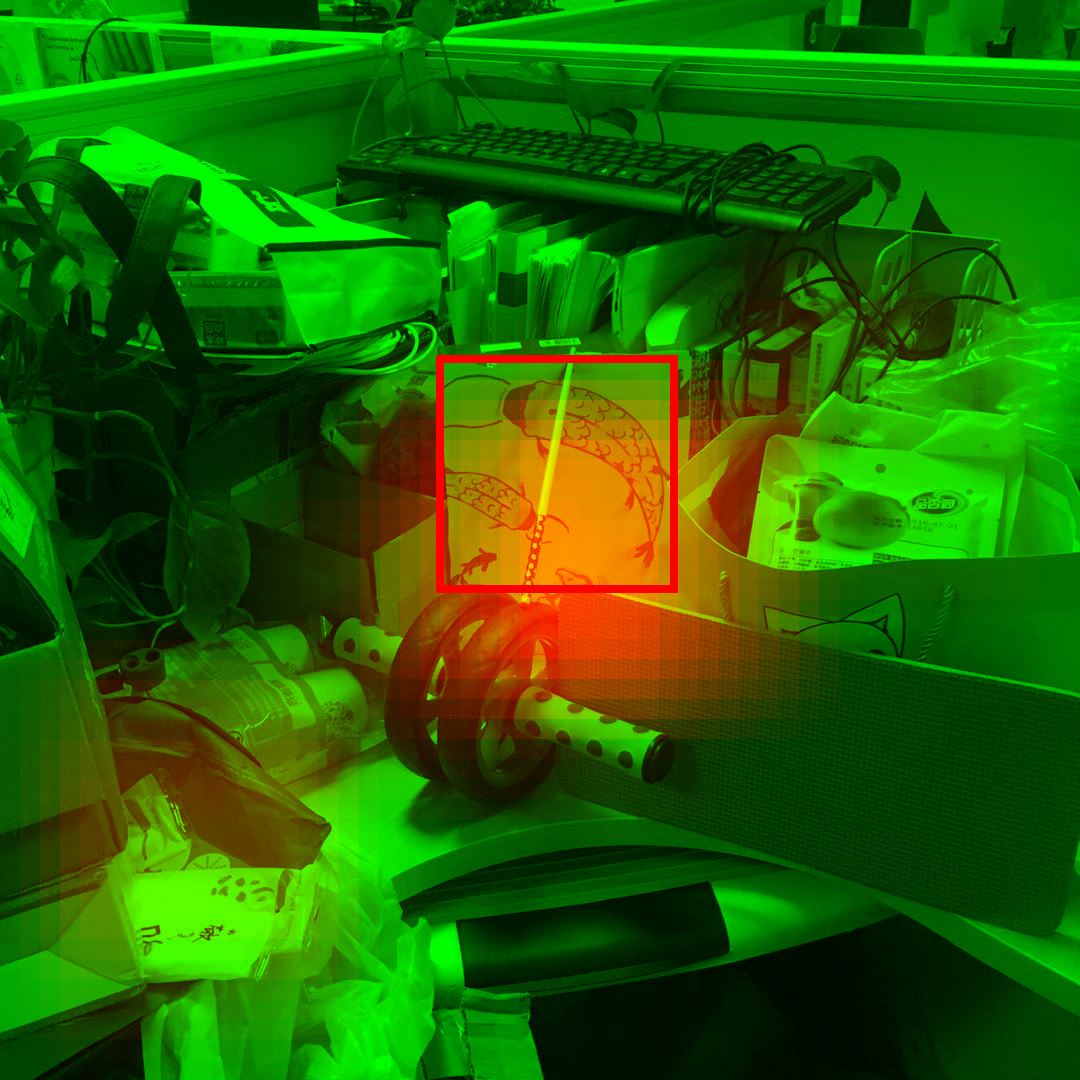}
    }
    \\
    \subfloat{
        \includegraphics[width=0.225\linewidth]{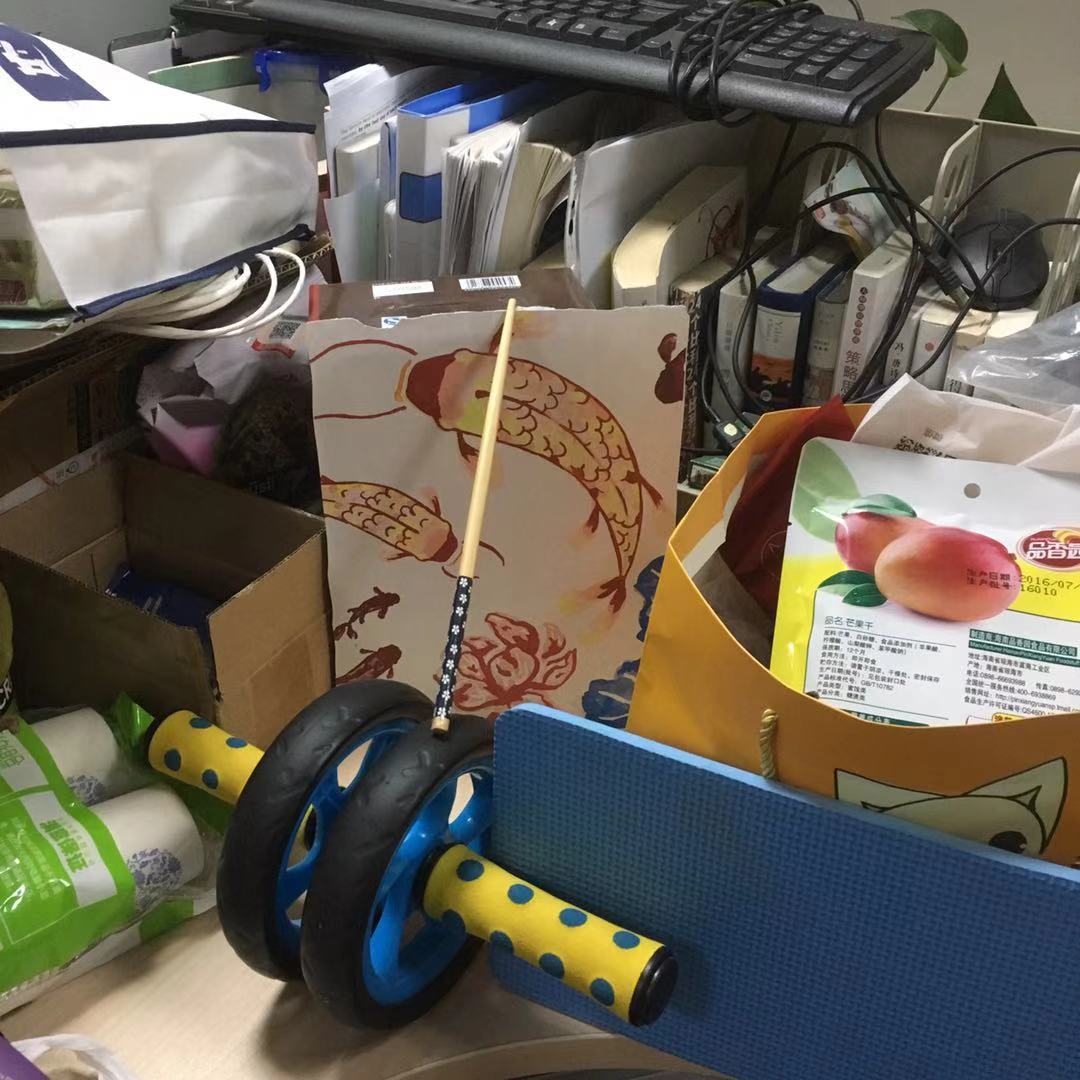}
    }
    \subfloat{
        \includegraphics[width=0.225\linewidth]{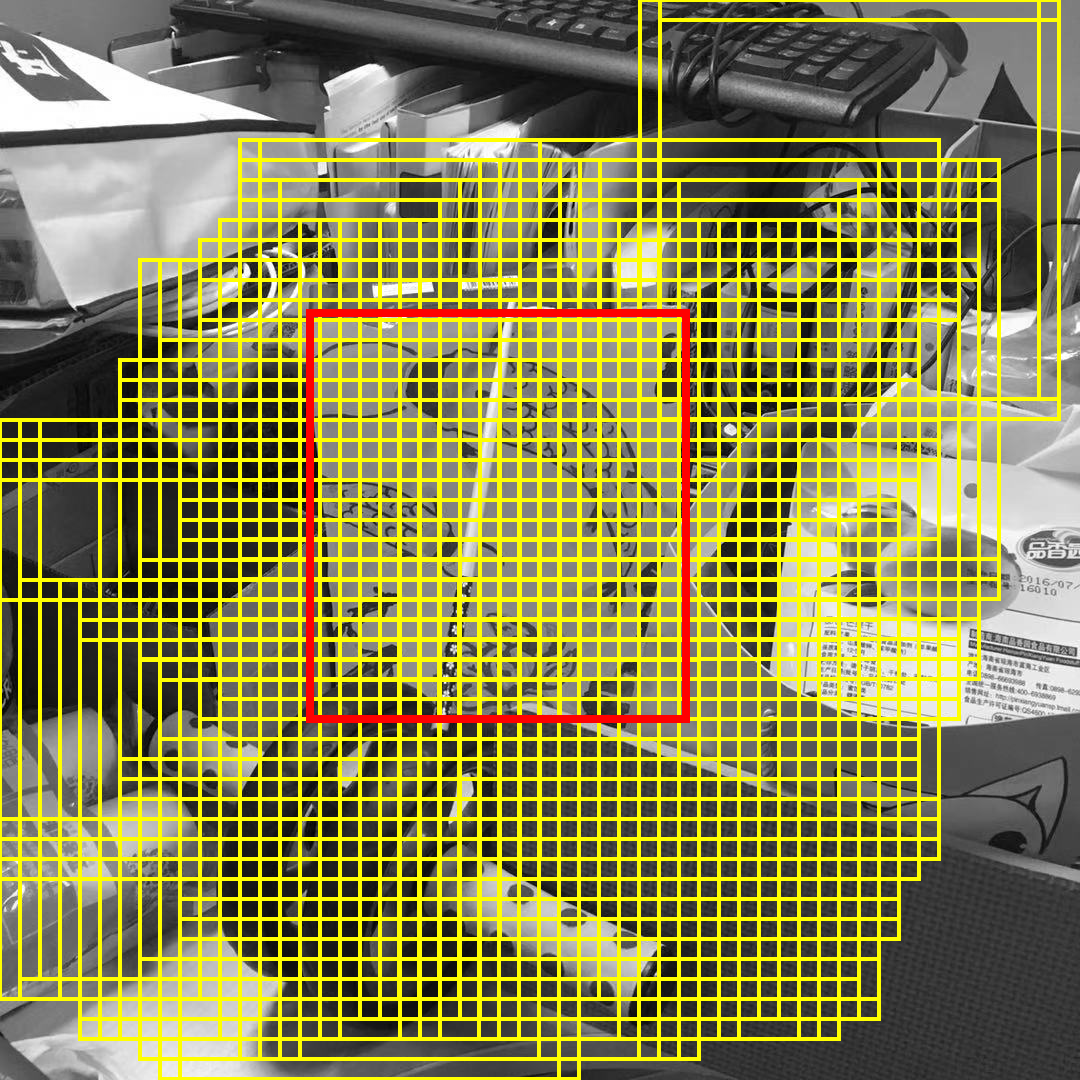}
    }
    \subfloat{
        \includegraphics[width=0.225\linewidth]{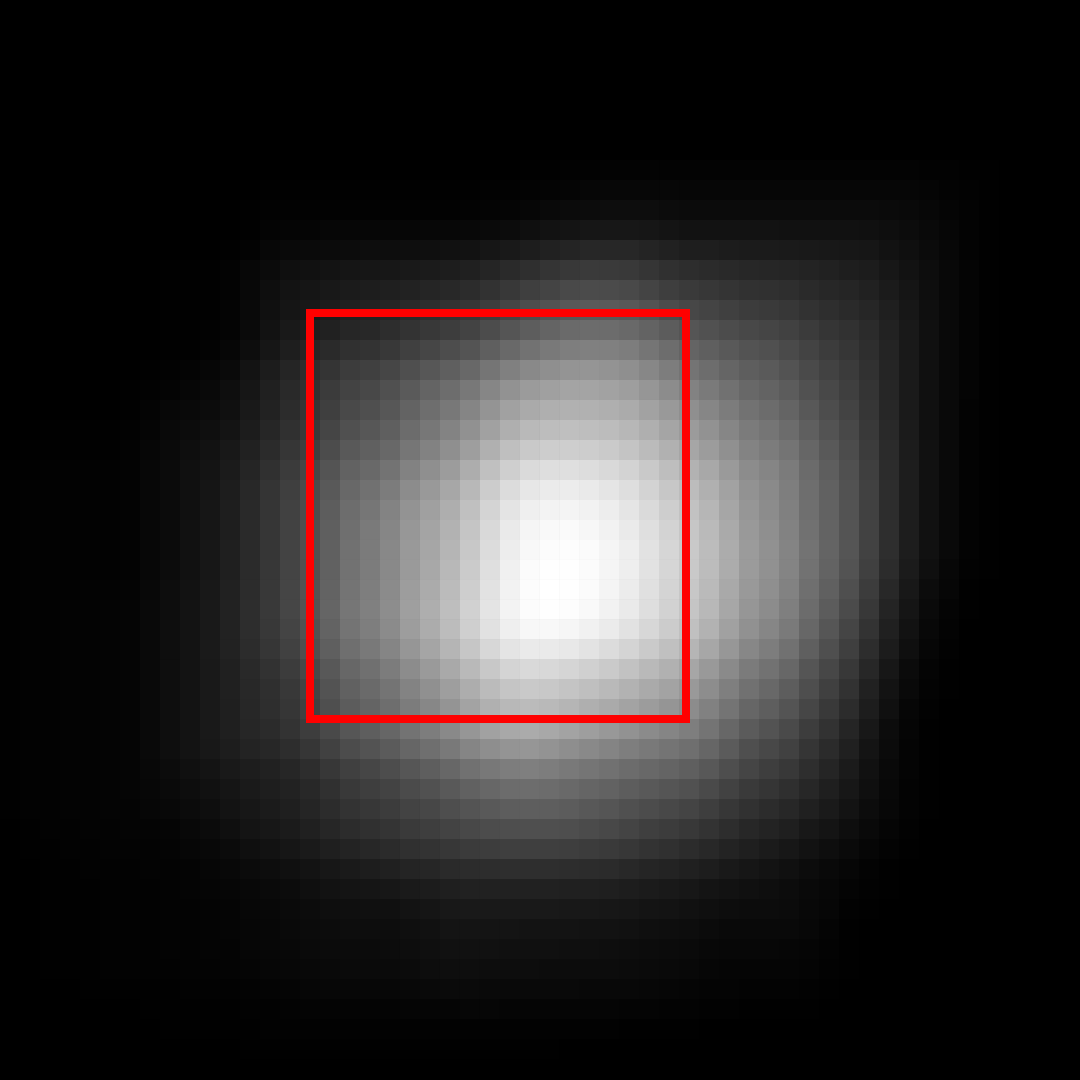}
    }
    \subfloat{
        \includegraphics[width=0.225\linewidth]{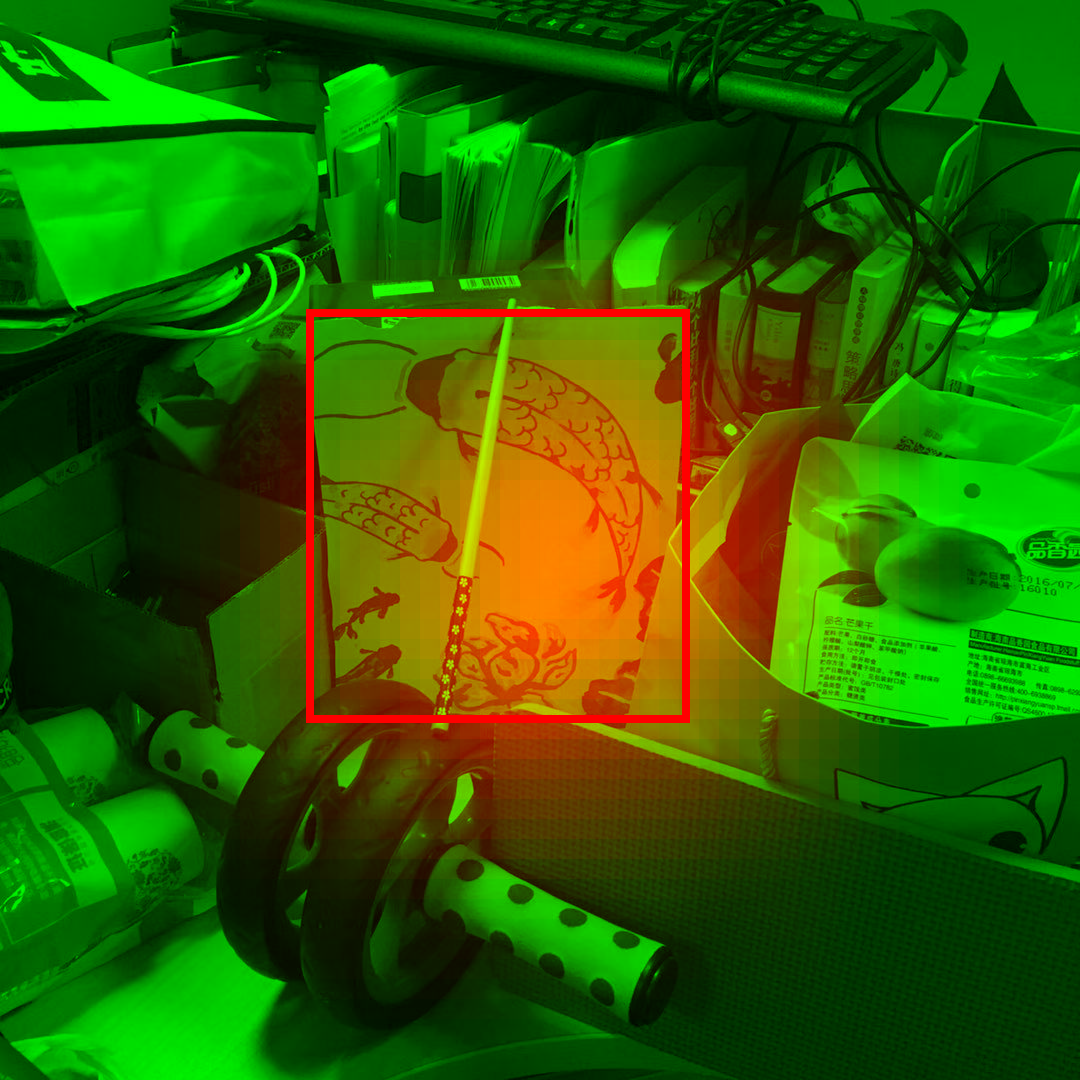}
    }
    \\
    \subfloat[Input\label{subfig:study2_input}]{
        \includegraphics[width=0.225\linewidth]{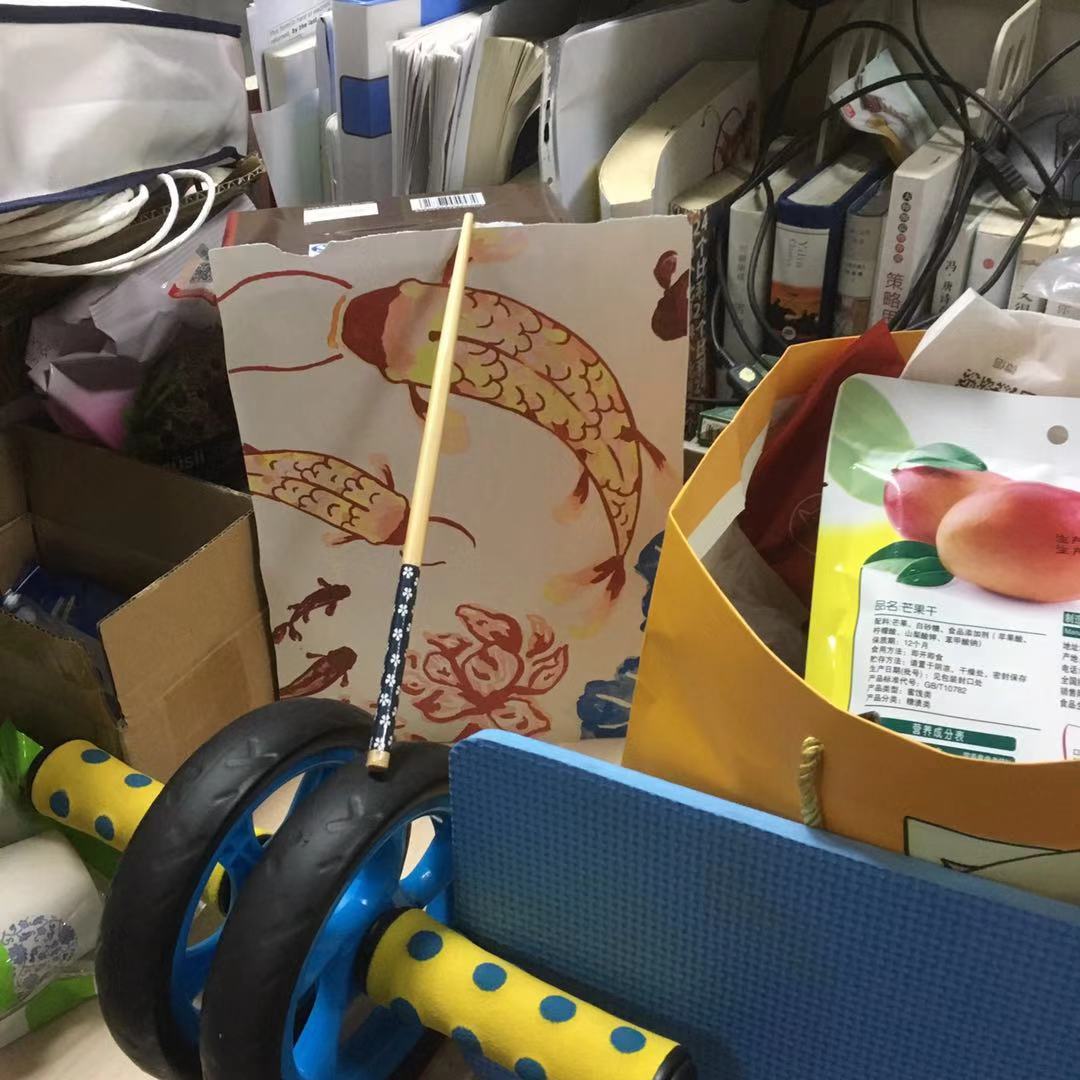}
    }
    \subfloat[Proposal\label{subfig:study2_proposal}]{
        \includegraphics[width=0.225\linewidth]{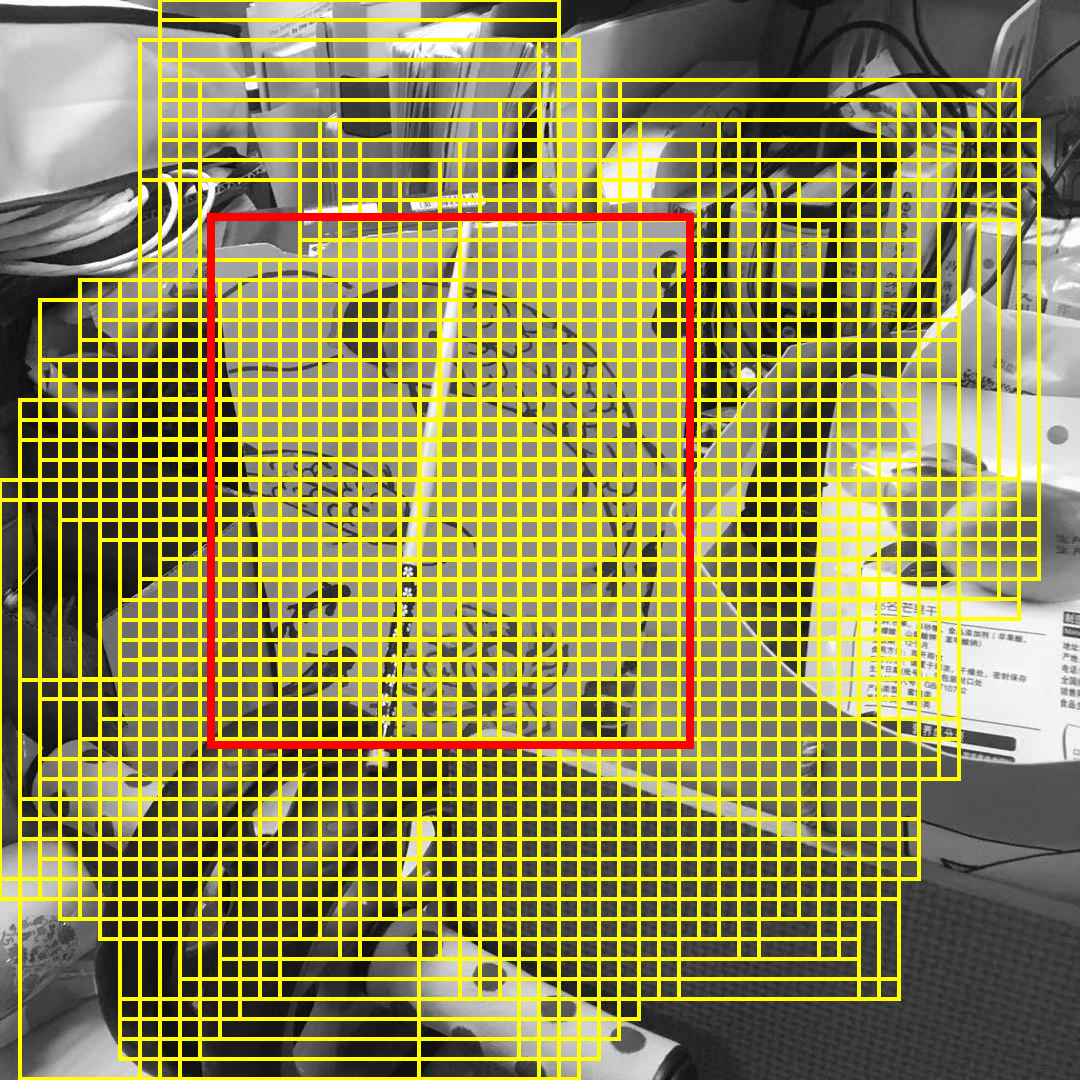}
    }
    \subfloat[Gray\label{subfig:study2_gray}]{
        \includegraphics[width=0.225\linewidth]{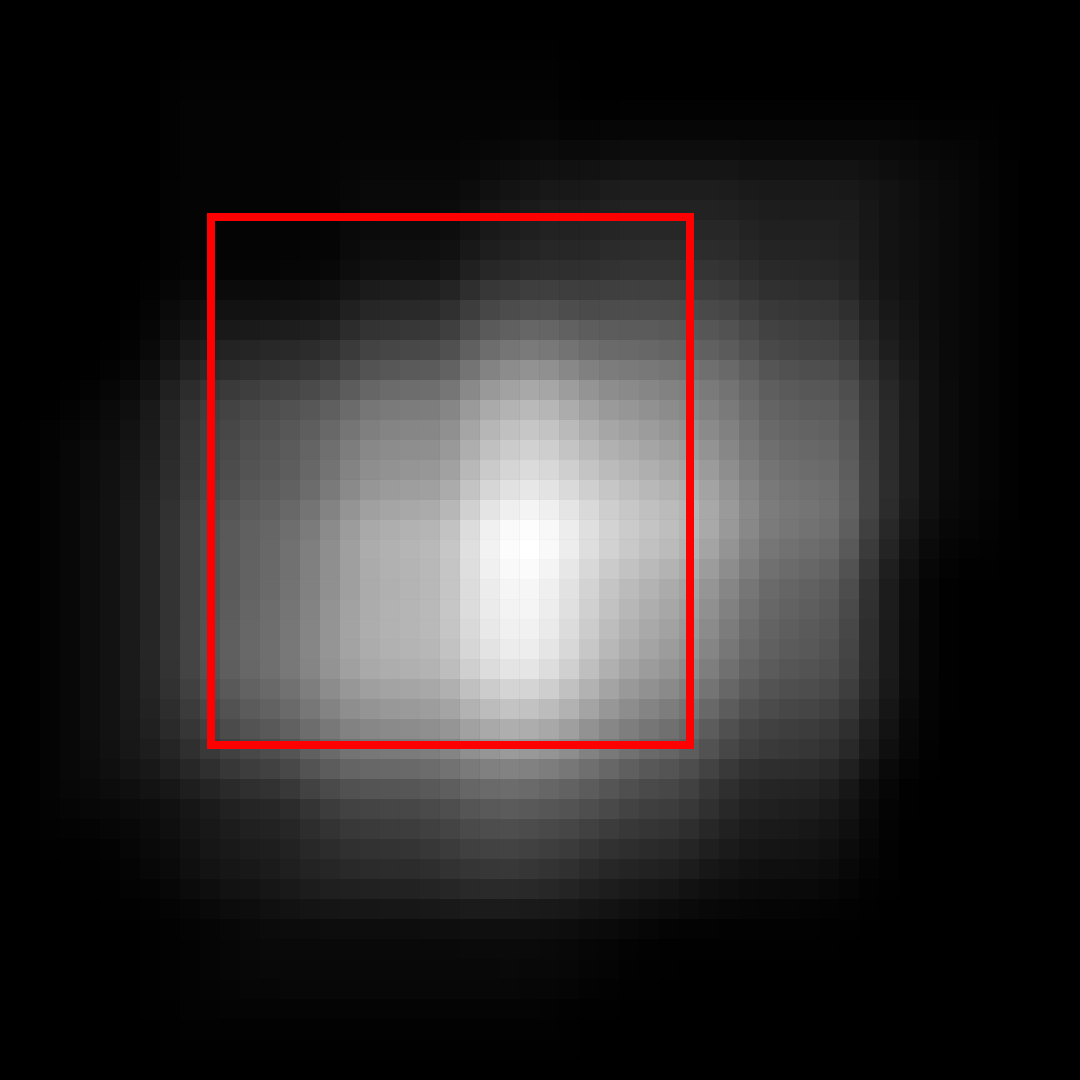}
    }
    \subfloat[Fused\label{subfig:study2_fused}]{
        \includegraphics[width=0.225\linewidth]{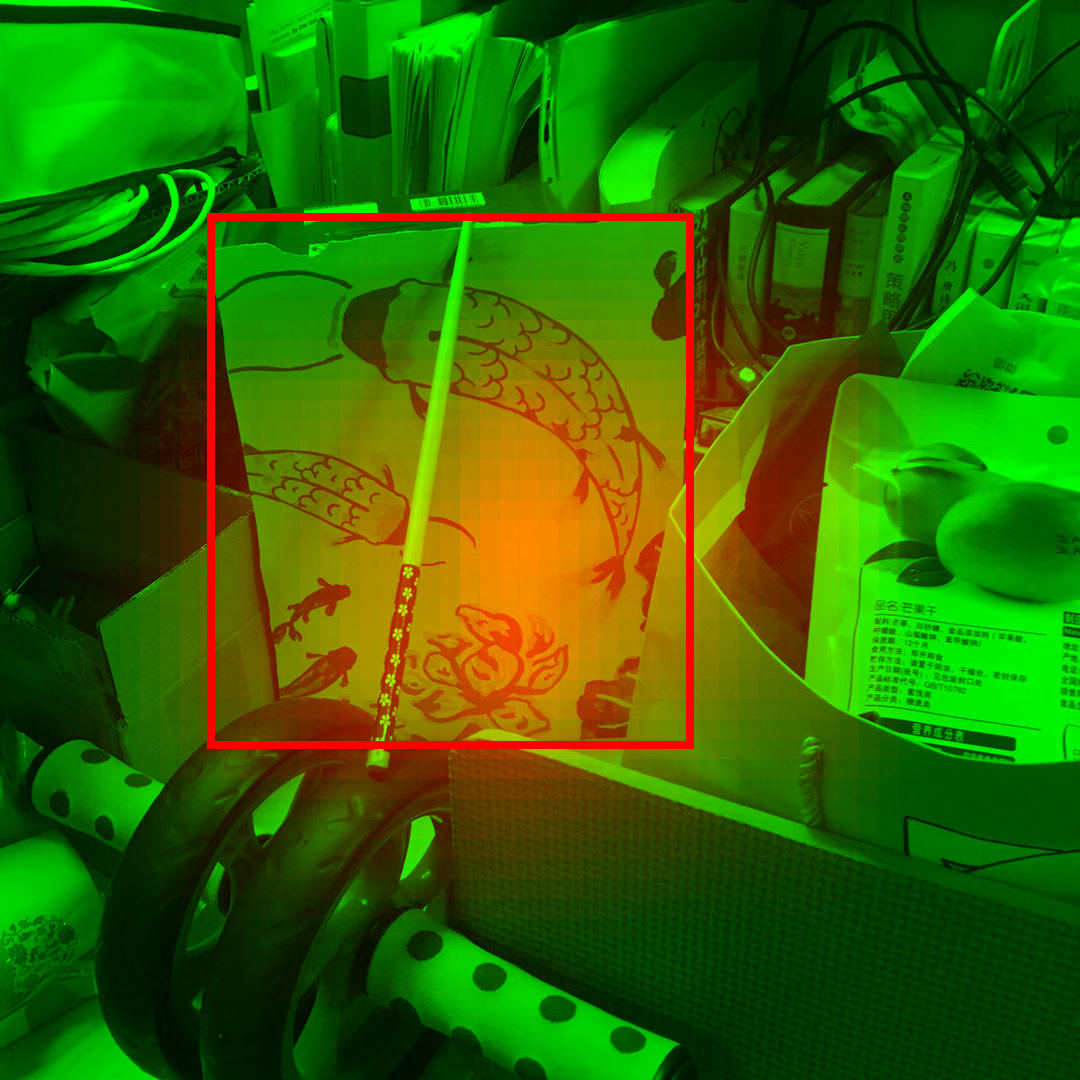}
    }
    \caption{Complex Artcode detection study in cluttered background, poor lighting.}
    \label{fig:study_2} 
\end{figure}

Two studies were conducted, both involving five images sequences (Figures~\ref{subfig:study1_input} and \ref{subfig:study2_input}) captured with a smartphone moving from far away to close proximity to an Artcode. 
The size of the Artcode gradually increases as the smartphone moves towards to the target, from top to bottom in the left-most column of the figures (Figures \ref{subfig:study1_input} and \ref{subfig:study2_input}).
Recognition is more challenging from further away. 
Apart from this, the two studies other settings differed as follows: 
The first study, Figure \ref{fig:study_1}, used a simple Artcode design, in good lighting, with an uncluttered scene, and an unoccluded Artcode. 
The second study, Figure \ref{fig:study_2}, involved a more difficult scenario, using a complex Artcode design, shaded lighting, a cluttered scene, and a partially occluded Artcode. 

Considering space limitations, and the focus of this paper, the technical details for building the Artcodes-detection machine-learning model are omitted.
Similarly, the details underlying the various elements in Figures \ref{fig:study_1} and \ref{fig:study_2} (including generation of the proposals and presence maps) are also omitted.
Interested readers are again referred to the literature for more information  \cite{meese2013codes, xu2017recognizing, xu2019artcode}.

\subsection{Results
\label{subsec:results}}


Figures \ref{fig:study_1} and \ref{fig:study_2} contain the content and results of the two studies.
The four columns in each figure, from left to right, are:
(a) the input images
(b) the Artcode proposals, annotated with yellow rectangles;
(c) the gray Artcode presence heat map; and 
(d) the fused image 
(created by combining the input image (a) with the heat map (c)). 
The red boxes indicate the ground-truth Artcodes. 

\begin{table}
    \begin{center}
        \caption{Decoding results for the images in Figures \ref{subfig:study1_input} and \ref{subfig:study2_input}.}
        \label{tab:decoding}
        \begin{tabular}{|c|c|c|c|c|c|}
        \hline
        \diagbox{Decoded}{Image} & 1st (top) & 2nd & 3rd & 4th & 5th (bottom) \\ 
        \hline\hline
        1st study (Figure \ref{fig:study_1}) & $\times$ & $\times$ & $\surd$ & $\surd$ & $\surd$ \\ 
        \hline
        2nd study (Figure \ref{fig:study_2}) & $\times$ & $\times$ & $\times$ & $\times$ & $\times$ \\ 
        \hline
        \end{tabular}
    \end{center}
\end{table} 

In addition to the presence detection results in Figures \ref{fig:study_1} and \ref{fig:study_2}, Table~\ref{tab:decoding} presents the decoding results (generated according to Artcode decoding procedures \cite{meese2013codes}).
Ticks and crosses in the table indicate whether the given image was successfully decoded or not, with ticks (``$\surd$'') indicating success; and 
crosses (``$\times$'') indicating failure.

It is clear that the detection proposals in both studies cover the actual marker areas
---
the penguins in Figure \ref{fig:study_1}, and the fish in Figure \ref{fig:study_2}
---
in all image sequences, with dense accumulation of the proposal rectangles centering around the target markers. 
This is further evidenced in the presence maps (gray and fused), where the marker areas are distinctly visible as heat spots (the bright areas in the 3rd and 4th columns of Figures \ref{fig:study_1} and \ref{fig:study_2}). 
The Artcode proposals in all five of the first study images center around the true Artcode areas, identified by the red boxes:
In the second study, in contrast, although the Artcode proposals cover the true Artcode areas, there are multiple proposals that are not around the actual target, especially for the images that were captured from a greater distance (in the top three rows of Figure~\ref{subfig:study2_input}).

The cluttered scene in the second study affects the detection, increasing the number of false positives: Many non-Artcode objects in this scene may look like Artcodes, with their generic visual features potentially causing the classifier to label them as Artcodes.
However, although redundant heat spots were generated, the actual target Artcodes are also identified: 
Figures \ref{subfig:study2_gray} and \ref{subfig:study2_fused} show multiple detections (indicated by heat spots), but one of them does contain the actual target Artcode. 
Heat spots in the presence maps can alert the user to the possible existence of access points to the metaverse, encouraging the user to come close for follow-up examination and identification. 

According to the decoding results (Table~\ref{tab:decoding}), 
the top two images in the first study (those captured from the furthest distance) could not be decoded, due to the low resolution and loss of details. 
The closer three input images in the first study, however, ware successfully identified and decoded,  opening up the ``hidden'' virtual worlds. 
This represents a simplified realistic interaction, where the users often come closer to a target after first getting the general impression (the hint or clue). 

The more complicated environment in the second study, including a more sophisticated Artcode, poorer lighting, clutter, and occlusion (with a chopstick in the way), resulted in none of the five images being successfully decoded.
This also represents a common, real-world situation, where the target image may be obscured from certain angles. 
In this case, the presence maps should motivate the user to get nearer, and to remove the obstruction, or to explore new viewing angles for better identification.
The explorative interaction process allowed by the proposed URF would enable various designs (e.g., design for serendipity \cite{andre2009discovery, danzico2010design}), and open up new interaction opportunities for connecting to the metaverse. 
\section{Discussion and implications
\label{sec:discussion}}

The two studies present a simplified and concrete implementation of the proposed framework, illustrating the key steps of detecting and identifying visual markers before decoding them, and accessing the metaverse.
Currently, implementing the proposed URF for all known visual markers may not be feasible
---
partly due to the ever-expanding set of such markers, and the regular emergence of new interaction devices.
However, this investigation using Artcodes as a representative marker provides evidence for the URF's applicability.
This paper, and the URF generally, can also serve as guidance for metaverse access point design,
using visual markers (especially in an unobtrusive but explorative manner). 
The proposed framework also includes a mixed interaction manner, combining physical movement and digital engagement
in an augmented physical world with ubiquitous connection access points. 

\section{Conclusion and future work
\label{sec:conclusion}}

In this paper, we have explored the problem of connecting with virtual worlds (or the metaverse) in an augmented physical world. 
We have presented a unified recognition framework (URF) consisting of three components for designing and implementing an explorative access point. 
A concrete implementation of this URF using Artcodes as access points was used to illustrate the process. 
An example of visual markers, Artcodes are both machine-readable and human-meaningful decorative patterns that represent the kind of access tool that will become increasingly commonplace in the future. 
The initial discovery of the presence of markers (indicated by a heat map) and the follow-up, closer inspection and detection were demonstrated by the two studies in the paper. 
The URF would enable the design of a kind of brokering system that can invoke appropriate recognition algorithms to deal with different types of access points, and may inspire interaction design in the metaverse age.
While this study used smartphones and Artcodes, our future work will include the investigation of other AR devices and other visual markers.

\section*{Acknowledgments}

\blackout{
This work is supported by the Natural Science Foundation of China (project no. 61872167).
The authors acknowledge the financial support from the Artificial Intelligence and Optimisation Research Group (AIOP), the Faculty of Science and Engineering (FoSE), the International Doctoral Innovation Centre, Ningbo Education Bureau, Ningbo Science and Technology Bureau, and the University of Nottingham.}

\IEEEtriggeratref{8}
\bibliographystyle{ieeetr}
\bibliography{references}

\end{document}